\documentclass[a4paper, 12pt, reqno, twoside]{amsart}
\usepackage{amsmath, amsthm, amssymb, enumerate}
\usepackage{fullpage, color, tikz}
\usepackage[colorlinks=true,urlcolor=cyan,linkcolor=blue,citecolor=magenta]{hyperref}
\usepackage{makecell}

\DeclareTextFontCommand{\emph}{\bfseries}

\theoremstyle{plain}
\newtheorem{theorem}{Theorem}[section]
\newtheorem{proposition}[theorem]{Proposition}
\newtheorem{lemma}[theorem]{Lemma}
\newtheorem{corollary}[theorem]{Corollary}
\newtheorem{conjecture}[theorem]{Conjecture}

\newtheorem{problem}[theorem]{Problem}

\theoremstyle{definition}

\newtheorem{remark}[theorem]{Remark}

\newtheorem{question}[theorem]{Question}

\newcommand{\inpro}[2]{\langle{#1},{#2}\rangle}

\newcommand{\R}{\mathbb{R}}
\newcommand{\PP}{\mathbb{P}}
\newcommand{\Sph}{\mathbb{S}}
\newcommand{\set}[2]{\left\{#1\ \middle\vert\ #2\right\}}
\newcommand{\setsize}[2]{\left\vert\left\{#1\ \middle\vert\ #2\right\}\right\vert}
\newcommand{\size}[1]{\left\vert#1\right\vert}
\newcommand{\spn}{\operatorname{span}}
\newcommand{\rank}{\operatorname{rank}}

\title{All-even sets in $\R^2$ and $\R^3$}

\author{Daniel Kalmanovich}
\address{Department of Computer Science, Ariel University, Israel} 
\email{danielk@ariel.ac.il}
\urladdr{https://danielkalmanovich.wixsite.com/website}

\author{Gabriel Nivasch}
\address{Department of Computer Science, Ariel University, Israel} 
\email{gabrieln@ariel.ac.il}
\urladdr{https://www.gabrielnivasch.org/home}

\begin{document}

\begin{abstract}
Call a finite nonempty point set $S\subset\R^d$ \emph{all-even} if for every direction $\ell \in \PP(\R^d)$, the number of hyperplanes perpendicular to $\ell$ that intersect $S$ is even. In 2026, Francisco Santos presented a question of Juan Benigno Seoane-Sep\'ulveda on the existence of all-even sets in dimensions $3$ and higher, together with examples of such sets in dimension $2$.

Call a set $S\in\R^d$ \emph{line-all-even} if for every direction $\ell \in \PP(\R^d)$ the number of lines parallel to $\ell$ that intersect $S$ is even. As Santos pointed out, every all-even set in $\R^3$ is line-all-even. We conjecture that there exist no line-all-even sets in $\R^3$ that affinely span $\R^3$

%In this paper we introduce a conjecture, that if true would answer Seoane-Sep\'ulveda's question in dimension $3$ in the negative. Our conjecture is as follows: For every finite point set $S\subset \R^3$ that affinely spans $\R^3$, there exists a direction $\ell \in \PP(\R^3)$ such that the number of lines parallel to $\ell$ that intersect $S$ is odd.

We show that our conjecture is equivalent to the statement that every planar all-even set satisfies a certain property, which we call \emph{height-rigidity}. This property has a natural interpretation in linear algebra terms. Furthermore, we present a hierarchy of properties of planar point sets that imply height-rigidity.

We construct several all-even sets in the plane, including some infinite families, and show that  all the sets we constructed are height-rigid. 

In addition, we prove that every all-even set in the plane that can be covered by six or fewer parallel lines is height-rigid.
\end{abstract}

\maketitle

\section{Introduction}\label{sec:intro}

The cardinality of the image of a finite point set under projection reflects the coincidences created by the projection direction. In the plane, this quantity is also the number of occupied lines in a parallel family. Questions about small projection images and about the range of possible image sizes have been studied from several perspectives. Skiena~\cite{Skie90} investigated the number of directions admitting a projection with at most a prescribed number of distinct images, while Charlton et al.~\cite{CharDDDMU12} studied consecutive ranges of projection cardinalities attainable by a single planar set. In this paper we consider a simultaneous parity condition: can every one-dimensional projection of a nonempty finite point set have even cardinality?

The question motivating this work, due to Juan Benigno Seoane-Sep\'ulveda and presented by Francisco Santos at an open problems session at the conference Combinatorics and Geometry in Mytilene 2026, is whether every nonempty finite point set in $\R^3$ has a one-dimensional projection of odd cardinality. We investigate this question through the geometric restrictions imposed by requiring every such projection to have even cardinality.

By a \emph{direction} in $\R^d$ we mean an element $\ell$ of the real projective space $\PP(\R^d)$, the set of all one-dimensional linear subspaces
of $\R^d$. Meaning, for every $u\in \R^d\setminus\{0\}$, we have
$$
\ell=\spn (u)=\set{\lambda u}{\lambda\in\R}\in\PP(\R^d).
$$
We say that $\ell$ is the direction \emph{determined} by $u$. Given two distinct points $p,q\in\R^d$, we denote $pq=q-p$. Each direction $\ell\in\PP(\R^d)$ corresponds to two opposite \emph{sub-directions} $u_1,u_2\in\Sph^{d-1}$ that are parallel to $\ell$ and opposite to each other, meaning $\|u_1\|=\|u_2\|=1$, $\spn(u_1)=\spn(u_2)=\ell$, and $u_2=-u_1$.

Let $S\subset \R^d$ be a non-empty finite set of points. Given a direction $\ell\in\PP(\R^d)$, choose an arbitrary $u\in\R^d\setminus\{0\}$ such that $\ell=\spn(u)$, and define
$$
m_S(\ell)=\setsize{\inpro{p}{u}}{p\in S} .
$$
This is a well-defined positive integer, since replacing $u$ by a non-zero scalar multiple of it does not change the number of distinct values. We may occasionally abuse notation and write $m_S(u)$. Geometrically, $m_S(\ell)$ is the number of affine hyperplanes perpendicular to $\ell$ that meet $S$. We call $S$ \emph{all-even} if $m_S(\ell)$ is even for all $\ell\in\PP(\R^d)$.

For the case $d=3$, a direction $\ell\in\mathbb P(\R^3)$ such that $m_S(\ell)$ is odd is called an \emph{odd plane direction} for $S$. We formulate Seoane-Sep\'ulveda's question as follows:

\begin{question}[Seoane-Sep\'ulveda]
Is it true that every finite nonempty set $S\subset\mathbb R^3$ has an odd plane direction?
\end{question}

We conjecture that the answer is positive. In other words:

\begin{conjecture}[weak conjecture]\label{conj:wpc}
There exists no all-even set in $\R^3$: For every finite nonempty set $S\subset\R^3$ there exists a direction $\ell\in\mathbb P(\R^3)$ with $m_S(\ell)$ odd.
\end{conjecture}

Clearly, if $|S|$ is odd, then a generic direction is an odd plane direction for $S$. Also, if $S$ is coplanar, then a normal direction to the plane containing $S$ is an odd plane direction for $S$. Furthermore, for $S$ to be all-even, it cannot be centrally symmetric: Suppose $S$ is centrally symmetric around the origin, and let $\pi$ be a plane containing the origin and at least one point of $S$. Then $S\setminus\pi$ consists of two symmetric parts, so an even number of additional planes parallel to $\pi$ are required to cover $S$. Hence, $m_S(\ell)$ is odd for the direction $\ell$ orthogonal to $\pi$.

Thus, any all-even set in $\R^3$, if it exists, must have even size, it must affinely span $\R^3$, and it cannot be centrally symmetric.

The same question can be asked in any dimension. Moreover, if the answer is yes in $\R^d$, then the answer is yes in $\R^k$ for every $k\geq d$. Indeed, let $S\subset \R^k$ be finite, and let $\pi:\R^k\to\R^d$ be a surjective linear map. By the assumed result in $\R^d$, applied to $\pi(S)$, there exists $u\neq 0$ such that $m_{\pi(S)}(u)$ is odd. Since $\pi$ is surjective, $\pi^*(u)\neq 0$, and $m_S(\pi^*(u))=m_{\pi(S)}(u)$, which is odd.

\begin{figure}[h!]
\centering
\begin{tikzpicture}[scale=2.2]

% Parameters
\def\n{15}
\def\r{1}

% Draw the regular 15-gon lightly
\draw[gray!45, thin]
  ({\r*cos(90)},{\r*sin(90)})
  \foreach \i in {1,...,14} {
    -- ({\r*cos(90+360*\i/\n)},{\r*sin(90+360*\i/\n)})
  } -- cycle;

% Draw the removed regular triangle
\draw[gray!60, dashed]
  ({\r*cos(90+360*0/\n)},{\r*sin(90+360*0/\n)})
  -- ({\r*cos(90+360*5/\n)},{\r*sin(90+360*5/\n)})
  -- ({\r*cos(90+360*10/\n)},{\r*sin(90+360*10/\n)})
  -- cycle;

% Kept vertices
\foreach \i in {1,2,3,4,6,7,8,9,11,12,13,14} {
  \fill ({\r*cos(90+360*\i/\n)},{\r*sin(90+360*\i/\n)})
        circle (1.3pt);
}

% Removed vertices
\foreach \i in {0,5,10} {
  \draw[thick, fill=white]
    ({\r*cos(90+360*\i/\n)},{\r*sin(90+360*\i/\n)})
    circle (2pt);
}

% Optional labels
\foreach \i in {0,...,14} {
  \node[font=\scriptsize]
    at ({1.15*\r*cos(90+360*\i/\n)},{1.15*\r*sin(90+360*\i/\n)})
    {$\i$};
}

\end{tikzpicture}
\caption{Juan Benigno Seoane-Sep\'ulveda's example of a planar all-even set: the vertices of a regular $15$-gon with the vertices $0,5,10$ of a regular triangle removed. For every direction, the set is covered by an even number of parallel lines.}
\label{fig:paco-example}
\end{figure}
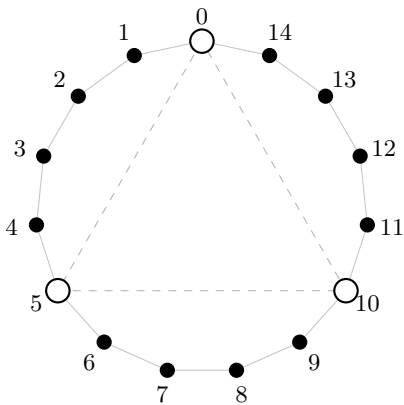

Seoane-Sep\'ulveda showed that the answer is no in $\R^2$, by constructing the all-even planar set shown in Figure~\ref{fig:paco-example}, which Santos presented at the same session.

One central contribution of our paper is a formulation that connects the three-dimensional problem with a rigidity property of planar configurations. We consider the stronger assertion that every finite set affinely spanning $\R^3$ has a direction with an odd number of occupied parallel lines, and prove that it is equivalent to height-rigidity of every planar all-even set. Here height-rigidity means that any assignment of heights preserving parallelism of secants is the restriction of an affine function. We develop three approaches to determine the validity of this assertion: propagation from small rigid subsets, a matrix formulation that relates all-evenness to the parity of directional ranks, and an analysis of configurations supported on few parallel lines. We construct several finite and infinite families admitting monotonically contagious triangles, and prove that every planar all-even set supported on at most six parallel lines is height-rigid, without any bound on the number of points on those lines.

\subsection*{Organization of the paper}

Sections~\ref{sec_lineAE} and~\ref{sec_rigidity} introduce our stronger conjecture and establish its equivalence to height-rigidity of planar all-even sets. Section~\ref{sec:rigidity_matrix} develops the algebraic formulation of height-rigidity, and Section~\ref{sec_contagion} introduces the hierarchy of contagion properties. Section~\ref{sec:constructions} presents constructions of planar all-even sets. Section~\ref{sec_singly} studies singly-determined directions, providing tools for the proof of the six-line theorem in Section~\ref{sec_6lines}. Section~\ref{sec:concluding} concludes with a discussion of some open problems. Appendix~\ref{app_180} contains the classification of non-collinear six-point sets with no singly-determined direction used in Section~\ref{sec_6lines}.

\section{line-all-even sets and a stronger conjecture}\label{sec_lineAE}

Let $S\subset\R^d$ be a nonempty finite point set. For a direction $\ell\in\mathbb P(\R^d)$, let
$$
\pi_\ell:\R^d\to \R^d/\ell
$$
be the quotient map, and define
$$
n_S(\ell)=|\pi_\ell(S)|.
$$
Equivalently, $n_S(\ell)$ is the number of affine lines parallel to $\ell$ that meet $S$. We say that $\ell$ is a \emph{secant} direction of $S$ if it is determined by a pair of points of $S$. A direction $\ell$ such that $n_S(\ell)$ is odd is called an \emph{odd line direction} for $S$. We call $S$ \emph{line-all-even} if it has no odd line direction. Clearly, if $|S|$ is odd, then every non-secant direction of $S$ is an odd line direction. Hence, every line-all-even set in $\R^d$ must have even size. Furthermore, every all-even set in $\R^2$, when considered as a subset of $\R^d$, is clearly a line-all-even set.

We conjecture the following in dimension $3$:

\begin{conjecture}[strong conjecture]\label{conj:slc}
Let $S\subset\R^3$ be a finite set that affinely spans $\R^3$. Then there exists a direction $\ell\in\mathbb P(\R^3)$ such
that $n_S(\ell)$ is odd.
\end{conjecture}

The strong conjecture implies the weak conjecture by the following observation, which Santos also presented at the same open problems session:

\begin{proposition}\label{prop:AE_is_LAE}
Let $S\subset\R^3$ be finite. If $S$ has an odd line direction, then it has an odd plane direction. In other words, if $S$ is all-even then it is line-all-even.
\end{proposition}

\begin{proof}
Let $\ell\in\mathbb P(\R^3)$ be a direction such that $n_S(\ell)$ is odd. Choose a non-zero vector $u\in \ell^\perp$ generically, so that distinct affine lines parallel to $\ell$ and meeting $S$ lie in distinct affine planes perpendicular to $u$. Then these planes are exactly the occupied $u$-levels of $S$, hence $m_S(u)=n_S(\ell)$, and $u$ determines an odd plane direction for $S$.
\end{proof}

Since any all-even set in $\R^3$ must affinely span $\R^3$, the strong conjecture indeed implies the weak conjecture. Hence, a proof of the strong conjecture would give a positive answer to Seoane-Sep\'ulveda's original question in $\R^3$.

As in the case of the weak conjecture, the validity of the strong conjecture can be asked in any dimension, and, if the answer is yes in $\R^d$, then the answer is yes in $\R^k$ for every $k\geq d\ge 2$. Indeed, let $S\subset \R^k$ be finite, of even size, that affinely spans $\R^k$. Choose a generic surjection $\pi:\R^k\to\R^d$, that is injective on $S$, and which creates no new parallelisms among the secants of $S$. Such a map exists by choosing $\ker(\pi)$ to avoid the finitely many secant lines and the finitely many $2$-planes spanned by pairs of nonparallel secants. By the assumed result in $\R^d$, applied to $\pi(S)$, there exists a direction $\ell$ such that $n_{\pi(S)}(\ell)$ is odd. Since $\size{S}$ is even, $\ell$ must be a secant direction determined by some $\pi(p),\pi(q)\in\pi(S)$. By the fact that no new parallelisms were created by $\pi$ we deduce that
$$
n_S(L)=n_{\pi(S)}(\ell)
$$
for $L=\spn(q-p)\subset\R^k$.

\section{Height-rigidity and the strong conjecture}\label{sec_rigidity}
The purpose of this section is to relate the strong conjecture to a rigidity property of planar all-even sets. Let us first note how the terminology from Sections~\ref{sec_lineAE} and~\ref{sec:intro} specializes to the plane. Let $A\subset\R^2$ be finite. Then $n_A(\ell)$ is the number of affine lines parallel to $\ell$ that meet $A$. In dimension two it is also equivalent to the hyperplane counting function. Indeed, if $L\in\mathbb P(\R^2)$ is a direction and $\ell=L^\perp$, then the affine hyperplanes perpendicular to $L$ are exactly the affine lines parallel to $\ell$. Hence
$$
m_A(L)=n_A(\ell).
$$
Thus, for finite subsets of $\R^2$, the all-even and line-all-even conditions coincide. We shall therefore simply say that $A$ is \emph{all-even} if $n_A(\ell)$ is even for every direction $\ell\in\mathbb P(\R^2)$. Notice that every all-even set has even cardinality, since for a generic direction $\ell$ no two points of $A$ lie on the same affine line parallel to $\ell$, and therefore $n_A(\ell)=|A|$.

\subsection{Admissible height functions}

Let $A\subset\R^2$ be finite. A \emph{height function} on $A$ is a
function
$$
z:A\to\R.
$$
It defines a \emph{lift} of $A$ to $\R^3$:
\begin{equation}\label{eq_lift}
\widetilde A_z=\set{(a,z(a))}{a\in A}.
\end{equation}

We say that $z$ is \emph{admissible} if planar parallel secants lift to parallel secants with the correct scaling. Explicitly, whenever $p,q,r,s\in A$, $p\neq q$, $r\neq s$, and
$$
q-p=\lambda(s-r)
$$
for some $\lambda\in\R\setminus\{0\}$, we require
$$
z(q)-z(p)=\lambda(z(s)-z(r)).
$$
Equivalently, if two planar secants of $A$ are parallel, then the corresponding lifted secants in $\widetilde A_z$ are parallel in $\R^3$. Note that we could have $q=s$, in which case, the condition requires that three different collinear points $p,q,r$ be lifted to collinear points.

The main property of admissible height functions is the following:

\begin{lemma}\label{lem_adm_preserves}
    Let $A\subset \R^2$ be a finite planar subset, let $z:A\to\R$ be an admissible height function, and let $\widetilde A_z$ be the lift of $A$ as in (\ref{eq_lift}). Then $A$ is all-even if and only if $\widetilde A_z$ is line-all-even.
\end{lemma}

\begin{proof}
    If $|A|$ is odd, then $A$ is not all-even and $\widetilde A_z$ is not line-all-even. Hence, suppose $|A|$ is even.
    
    Suppose first that $A$ has an odd line direction $\ell\in\PP(\R^2)$. Then $\ell$ must be a secant direction of $A$, determined by two points $p,q\in A$. Let $\hat\ell$ be the direction determined by the lifted points $(p,z(p))$, $(q, z(q))$. Since the height function $z$ preserves collinearities, we have $n_A(\ell) = n_{\widetilde A_z}(\hat\ell)$. Hence, $\widetilde A_z$ has an odd line direction as well.

    Conversely, suppose $\widetilde A_z$ has an odd line direction $\hat\ell\in\PP(\R^3)$. Again, $\hat\ell$ must be a secant direction of $\widetilde A_z$, determined by two points $(p,z(p))$, $(q, z(q))$. Let $\hat L_1,\ldots,\hat L_k$ be the lines with direction $\hat\ell$ covering $\widetilde A_z$, where $k$ is odd. The orthogonal projection of $\R^3$ into $\R^2$ cannot make any two of lines $\hat L_i$, $\hat L_j$ coincide: Points $(r,z(r))\in \hat L_i$, $(s,z(s))\in \hat L_j$ on these lines would satisfy $pq\parallel rs$, which would imply that $z$ is not admissible. Hence, letting $\ell$ be the direction determined by $pq$, we have $n_A(\ell)=k$, so $A$ has an odd direction.
\end{proof}

We call $A$ \emph{height-rigid} if every admissible height function on $A$ is affine on $A$, that is, if every admissible $z:A\to\R$ has the form
$$
z(x,y)=\alpha x+\beta y+\gamma
$$
on the points of $A$.

\subsection{The rigidity reduction}

We first record the generic projection fact that will be used in the reduction.

\begin{lemma}\label{lem_surj}
Let $S\subset\R^3$ be finite. There exists a surjective linear map
$\pi:\R^3\to\R^2$ such that $\pi$ is injective on $S$ and creates no new
parallelisms among secants of $S$. Namely, whenever $p,q,r,s\in S$, $p\neq q$,
$r\neq s$, and
$$
\pi(q)-\pi(p)\parallel \pi(s)-\pi(r),
$$
we already have
$$
q-p\parallel s-r.
$$
\end{lemma}

\begin{proof}
Choose the kernel of $\pi$ to be a line $K\subset\R^3$. We need to avoid
finitely many bad choices of $K$.

First, $K$ should contain no non-zero secant vector $q-p$ with $p,q\in S$.
This guarantees that $\pi$ is injective on $S$.

Second, if $q-p$ and $s-r$ are non-parallel secant vectors of $S$, then $K$
should not lie in the plane $\spn(q-p,s-r)$. There are only
finitely many such planes, so a generic line $K$ avoids all of them.

Now suppose that
$$
\pi(q)-\pi(p)\parallel \pi(s)-\pi(r).
$$
Then some non-trivial linear combination of $q-p$ and $s-r$ lies in $K$.
If $q-p$ and $s-r$ were not parallel, then $K$ would lie in
$\operatorname{span}(q-p,s-r)$, contrary to the choice of $K$. Hence
$q-p\parallel s-r$.
\end{proof}

\begin{theorem}[Rigidity reduction]\label{thm:slc_equiv_AEisHR}
The strong conjecture in $\R^3$ (Conjecture \ref{conj:slc}) is equivalent to the statement: every all-even set $A\subset\R^2$ is height-rigid.
\end{theorem}

\begin{proof}
Suppose first that there exists an all-even set $A\subset \R^2$ that is not height-rigid. Then $A$ has an admissible height function $z$ such that $\widetilde A_z$ affinely spans $\R^3$. But Lemma \ref{lem_adm_preserves} implies that $\widetilde A_z$ is line-all-even, contradicting Conjecture \ref{conj:slc}.

For the other direction, suppose that every all-even set in $\R^2$ is height-rigid. Suppose, contrary to Conjecture \ref{conj:slc}, that there exists a line-all-even set $\hat A\subset \R^3$ that affinely spans $\R^3$. Let $\pi$ be the map guaranteed by Lemma \ref{lem_surj}. By applying a shear parallel to the $xy$-plane, we can assume without loss of generality that that $\pi$ is in fact an orthogonal projection (since affinities preserve parallelisms as well as the line-all-even property). In other words, $\pi$ maps every point $(a_x,a_y, a_z)\in \hat A$ into $(a_x,a_y)\in A\subset \R^2$. But then, the height function $z:A\to\R$ that provides back the $z$-coordinate $a_z$ of every point $(a_x,a_y)\in A$, is an admissible height function, contradicting the assumption that $A$ is height-rigid.
\end{proof}

\section{The rigidity matrix: an algebraic formulation}\label{sec:rigidity_matrix}

The definition of height-rigidity can be expressed in algebraic terms. Let
$$
A=\{a_1,\ldots,a_n\}\subset\R^2
$$
be finite and non-collinear, and let $\mathbf e_1,\ldots,\mathbf e_n$ denote the standard basis of $\R^n$. The \emph{height-rigidity matrix} $R(A)$ has a column for each point of $A$, and has a row for each parallel pair of secants. That is, whenever we have two parallel secants $\{i,j\}$ and $\{k,l\}$ satisfying
$$
a_j-a_i=\lambda(a_l-a_k),
$$
we include the row
$$
\mathbf e_j-\mathbf e_i
-\lambda\mathbf e_l+\lambda\mathbf e_k.
$$
The secants may share a point. Their orientations and order affect the row only by a non-zero scalar. Repeated rows are retained.

Considering a height function on $A$ as a vector $\mathbf z=(z_1,\dots,z_n)\in\R^n$, the equation
$$
R(A)\cdot\mathbf z=0
$$
is precisely the condition that $\mathbf z$ be admissible. In other words, the kernel of $R(A)$ is the space of admissible height functions. Writing $x_A,y_A$ for the coordinate vectors of $A$, the kernel always contains the $3$-dimensional space of affine heights
$$
\spn\left(\mathbf 1, x_A, y_A\right)\subseteq \ker R(A).
$$
Thus, we have
\begin{equation}\label{eq:HR_equiv_dimker3}
A\text{ is height-rigid}
\quad\Longleftrightarrow\quad
\dim\ker R(A)=3    
\end{equation}
and, more generally, the dimension of the space of admissible heights modulo affine heights is $n-3-\rank R(A)$.

Height-rigidity is a special case of affine rigidity in the sense of
Gortler et al.~\cite{GortGLT13}. Associate to $A$ the hypergraph whose
hyperedges are its collinear triples and its non-collinear four-point
subsets formed by two parallel secants. A height function is admissible
precisely when its restriction to each hyperedge is the restriction of
an affine function, which may depend on the hyperedge. The affine
dependencies on these hyperedges span the row space of $R(A)$;
relations supported on four collinear points are already generated by
the triple relations. Thus $R(A)$ has the same row space as a strong
affinity matrix for this realised hypergraph, and the criterion
$\dim\ker R(A)=3$ is the corresponding affine-rigidity criterion
\cite[Theorem~1]{GortGLT13}.

Having characterized height-rigidity by the rank of $R(A)$, we now express the all-even condition in terms of the ranks of its row blocks, grouped by secant direction. For each secant direction $\ell$, let
$$
\mathcal E_\ell(A)
=\set{\{i,j\}\in\tbinom{[n]}2}{a_j-a_i\in\ell}.
$$
These \emph{secant-direction blocks} partition $\binom{[n]}2$. Let $R_\ell(A)$ consist of the rows of $R(A)$ generated by pairs of secants in $\mathcal E_\ell(A)$. Thus, after permuting its rows, $R(A)$ has the block form
$$
R(A)=
\begin{pmatrix}
R_{\ell_1}(A)\\
\vdots\\
R_{\ell_m}(A)
\end{pmatrix},
$$
where $\ell_1,\ldots,\ell_m$ are the secant directions of $A$. A singleton secant-direction block contributes no rows.

\begin{proposition}[Directional rank]\label{prop:directional_rank}
For every secant direction $\ell$,
\[
\dim\ker R_\ell(A)=n_A(\ell)+1,
\qquad
\rank R_\ell(A)=n-n_A(\ell)-1.
\]
\end{proposition}

\begin{proof}
Let $L_1,\ldots,L_c$ be the occupied lines parallel to $\ell$, where $c=n_A(\ell)$, and choose a linear functional $\varphi:\R^2\to\R$ that does not vanish on $\ell$. The equations of $R_\ell(A)$ hold exactly when
\[
z_i=\sigma\varphi(a_i)+\gamma_r
\qquad\text{for }a_i\in L_r,
\]
with one common slope $\sigma$ and one intercept $\gamma_r$ per occupied line. Since some $L_r$ contains two points, these $c+1$ parameters are uniquely determined by $\mathbf z$. The nullity is therefore $c+1$, and the rank formula follows.
\end{proof}

\begin{corollary}\label{cor:all_even_directional_ranks}
The set $A$ is all-even if and only if $n$ is even and $\rank R_\ell(A)$ is odd for every secant direction $\ell$.
\end{corollary}

\begin{proof}
Apply Proposition~\ref{prop:directional_rank} to secant directions. Every other direction has $n$ occupied lines.
\end{proof}

The rigidity matrix records enough information to determine whether $A$ is all-even:

\begin{proposition}[Combinatorial reconstruction]\label{prop:rigidity_combinatorial_reconstruction}
The row space of $R(A)$ determines, up to relabeling the points, the secant-direction blocks and the occupied-line partition for each block. In particular, it determines the counts $n_A(\ell)$ and whether $A$ is all-even.
\end{proposition}

\begin{proof}
Put $W=\operatorname{row}R(A)$. For distinct secants $\{i,j\}$ and $\{k,l\}$ and any $\lambda\ne0$, we have
$$
a_j-a_i=\lambda(a_l-a_k)
\quad\Longleftrightarrow\quad
\mathbf e_j-\mathbf e_i-\lambda(\mathbf e_l-\mathbf e_k)\in W.
$$
The forward implication follows from the definition of $R(A)$. The reverse follows because every vector in $W$ annihilates $x_A$ and $y_A$.

This criterion recovers the blocks $\mathcal E_\ell(A)$. For each block, the connected components of the graph $\left([n],\mathcal E_\ell(A)\right)$, including isolated vertices, are precisely the occupied lines parallel to $\ell$. Their numbers and sizes give the claimed data; non-secant directions have $n$ occupied lines.
\end{proof}

\begin{remark}\label{rem:rigidity_overlapping_blocks}
Different direction blocks may contain proportional rows. For example, a parallelogram relation
\[
a_p+a_q=a_r+a_s
\]
gives the same row, up to scaling, for both $a_pa_r\parallel a_qa_s$ and $a_pa_s\parallel a_qa_r$. Thus the blocks classify the generating secants, not disjoint sets of numerical row values.
\end{remark}

\begin{proposition}[Affine reconstruction]\label{prop:rigidity_affine_reconstruction}
If $A$ is height-rigid, the row space of $R(A)$ determines $A$ up to affine equivalence and relabelling.
\end{proposition}

\begin{proof}
Choose a basis $\{\mathbf1,x,y\}$ of $\ker R(A)$. Since $\{\mathbf1,x_A,y_A\}$ is another basis, the points $(x_i,y_i)$ are obtained from the original points $a_i$ by an invertible affine transformation.
\end{proof}

Both reconstruction results survive row reduction. The height-rigidity assumption is essential for the second: when $\dim\ker R(A)>3$, the original subspace of affine heights is not, in general, determined by the row space alone.

\begin{remark}\label{rem:rigidity_strong_conjecture}
By the rigidity reduction, the strong conjecture is equivalent to $\rank R(A)=n-3$ for every non-collinear planar all-even set $A$. Corollary~\ref{cor:all_even_directional_ranks} translates all-evenness into parity conditions on the directional ranks. These ranks cannot simply be added, since the directional row spaces may intersect. The remaining problem is to establish the required global rank.
\end{remark}

\section{Contagion strategies}\label{sec_contagion}
We describe several propagation rules that provide sufficient conditions for height-rigidity. Throughout this section, $A\subset\R^2$ is finite and non-collinear, and $n=|A|$. The strategies apply whether or not $A$ is all-even.

\subsection{Contagious triangles}\label{subsec:cont_tri}
Let $T=\{a,b,c\}\subseteq A$ be a non-collinear triple. Start with the partition of $A$ in which $T$ is one class and the remaining points are singleton sets. Repeatedly merge the classes containing $x,y$ whenever $xy\parallel pq$, for some distinct points $p,q$ belonging to the same class. The secants $xy,pq$ may share an endpoint. The process terminates at a partition independent of the order of the merges. We call $T$ a \emph{contagious triangle} if the terminal partition consists of one class.

\begin{lemma}
    If $A$ has a contagious triangle, then $A$ is height-rigid.
\end{lemma}
\begin{proof}
Let $z:A\to\R$ be admissible, let $f$ be the affine function agreeing with $z$ on $T$, and put $h=z-f$. Then $h$ is admissible and vanishes on $T$. If $h(p)=h(q)$ and $y-x=\lambda(q-p)$, admissibility gives
$$
h(y)-h(x)=\lambda\left(h(q)-h(p)\right)=0.
$$
Thus $h$ is constant on every class throughout the process. If $T$ is contagious, $h$ vanishes on all of $A$, so $z=f$.    
\end{proof}

A triangle $T=\{a_1,a_2,a_3\}$ is called \emph{monotonically contagious} if the remaining points can be ordered as $a_4,\ldots,a_n$ so that, for every $r>3$,
$$
a_r-a_k\parallel a_j-a_i
\qquad\text{for some }i,j,k<r,\quad i\ne j.
$$
Equivalently, starting with $U=T$, one can repeatedly add a point $y\in A\setminus U$ for which $xy\parallel pq$, with $x,p,q\in U$ and $p\ne q$. This only uses merges that attach a singleton to the class containing $T$. Consequently, every monotonically contagious triangle is contagious.

\subsection{Contagion of directions}
For a finite set $X\subset\R^2$, let
$$
D(X)=\set{\spn(q-p)}{p,q\in X,\ p\ne q}
$$
be the set of all directions determined by $X$. Fix an integer $k\ge3$, and let
$$
\mathcal R_k(A)=
\set{S\subseteq A}{S\text{ is non-collinear and height-rigid},\ |S|\le k}.
$$
Given a pair $L_0=\{d_1,d_2\}\subseteq D(A)$ of distinct directions, define
$$
U_0=L_0,\qquad
U_{t+1}
=
U_t\cup
\bigcup_{\substack{S\in\mathcal R_k(A)\\
                   |D(S)\cap U_t|\ge2}}
D(S).
$$
The \emph{$k$-closure} of $L_0$ is
\begin{equation}\label{eq_def_closure}
\operatorname{cl}_k(L_0)=\bigcup_{t\ge0}U_t.
\end{equation}
We call $L_0$ \emph{$k$-contagious} if $\operatorname{cl}_k(L_0)=D(A)$, and write $P_k(A)$ for the assertion that $A$ has such a pair. Clearly, we have $P_k(A)\Longrightarrow P_{k+1}(A)$.

\begin{proposition}
    If $P_k(A)$ holds for some $k$, then $A$ is height-rigid.
\end{proposition}
\begin{proof}
    Let $z$ be admissible. Subtract a linear function whose slopes in $d_1,d_2$ agree with those of $z$, obtaining an admissible function $h$ with zero slope in both directions. Such a linear function exists because $d_1,d_2$ are distinct. Suppose $S\in\mathcal R_k(A)$ has two directions already known to have zero slope. Since $S$ is height-rigid, $h|_S$ is affine. Its linear part vanishes on two distinct directions, so $h|_S$ is constant. Hence every direction in $D(S)$ has zero slope. Induction proves this for every direction in the closure. If the pair is $k$-contagious, $h$ is constant on $A$, so $z$ is affine.
\end{proof}

For $k=3$, the set $\mathcal R_3(A)$ consists of all triples of non-collinear points; hence the rule simply adds the third side direction of a non-collinear triangle whenever its other two directions are infected. For $k=4$, the additional sets in $\mathcal R_4(A)$ are characterized as follows.

\begin{lemma}
    A non-collinear four-point set is height-rigid if and only if it contains three collinear points or two disjoint parallel secants.
\end{lemma}
\begin{proof}
    Its rigidity matrix has rank at most one. It has rank one exactly when it contains a non-zero row, which occurs precisely when two distinct secants are parallel. These either share an endpoint, giving three collinear points, or are disjoint.
\end{proof}

\subsection{The hierarchy of contagion strategies}
Write $M(A)$ for the existence of a monotonically contagious triangle in $A$, $C(A)$ for the existence of a contagious triangle in $A$, and recall from Section~\ref{subsec:cont_tri} that we have $M(A)\Longrightarrow C(A)$. 

Next, we show that $C(A)$ is in fact equivalent to $P_3(A)$.

\begin{proposition}
    The set $A$ has a contagious triangle if and only if $P_3(A)$ holds. Moreover, any two side directions of a contagious triangle form a $3$-contagious pair.
\end{proposition}
\begin{proof}
    For the first direction, let $T$ be a contagious triangle, let $d_1,d_2$ be any two side directions of $T$, and let
    $$
    U=\operatorname{cl}_3(\{d_1,d_2\})
    $$
    be the final set of directions produced by the $3$-contagion process, according to (\ref{eq_def_closure}). We will show that $U=D(A)$. Define
    $$
    x\equiv_U y
    \quad\Longleftrightarrow\quad
    x=y\ \text{or}\ \spn(y-x)\in U.
    $$
    This is an equivalence relation. Reflexivity and symmetry are immediate. For transitivity, suppose $x,y,z$ are distinct and the directions of $xy,yz$ belong to $U$. If the points are collinear, $xz$ has the same direction; otherwise, the definition of $3$-closure ensures that the third side direction also belongs to $U$.
    We claim that, throughout the triangle-contagion process induced by $T$, every class of the partition is contained in a single $\equiv_U$-class. This follows by induction. Initially, the vertices of $T$ are $\equiv_U$-equivalent because $d_1,d_2\in U$, and all other classes are singletons. Suppose the classes containing $x$ and $y$ are merged because $xy\parallel pq$, where distinct $p,q$ lie in a common class. By the induction hypothesis, $p\equiv_U q$, so the direction of $pq$, and hence of $xy$, belongs to $U$. Thus $x\equiv_U y$, meaning that the two classes being merged lie in the same $\equiv_U$-class. Since $T$ is contagious by assumption, the procedure ends with the single class $A$, i.e., all points are $\equiv_U$-equivalent, giving $U=D(A)$, as desired.
    
    For the second direction, suppose $\{d_1,d_2\}$ is $3$-contagious. Since $A$ is non-collinear, $A$ determines at least three directions. Hence there is a first addition, using a triangle $T$ with side directions $d_1,d_2$. Perform the triangle-contagion process starting with $T$, and let $\mathcal P$ be the resulting partition obtained. We will show that $\mathcal P=\{A\}$. Put
    $$
    V=\bigcup_{B\in\mathcal P}D(B).
    $$
    Every secant $xy$ whose direction belongs to $V$ has both endpoints in one class: there are distinct $p,q$ in a common part with $xy\parallel pq$, so otherwise another merge would be possible. Consequently, if two side directions of a non-collinear triangle belong to $V$, all three vertices lie in one class, and its third side direction also belongs to $V$. Since $d_1,d_2\in D(T)\subseteq V$, every direction added from $\{d_1,d_2\}$ belongs to $V$. Thus
    $$
    D(A)=\operatorname{cl}_3(\{d_1,d_2\})\subseteq V.
    $$
    Every pair of points therefore lies in one part of $\mathcal P$, so $\mathcal P=\{A\}$, as claimed.
\end{proof}

To summarize the hierarchy of the contagion properties, we have:
$$
M(A)\Longrightarrow C(A)\Longleftrightarrow P_3(A)
\Longrightarrow P_4(A)\Longrightarrow\cdots
\Longrightarrow P_n(A)\Longleftrightarrow
A\text{ is height-rigid}.
$$

As we will show in Section \ref{sec:constructions} below, all the all-even sets we know of satisfy the strongest property $M$. In the next subsection we show that the reverse implications do not hold in general.

\subsection{Separation between contagion properties}\label{subsec:sep_contagion_properties}

The following lemmata exhibit constructions of non-all-even sets that satisfy some of the weaker contagion properties but not the stronger ones. It is an open question whether there exists an all-even set that does not satisfy the strongest property $M$.

\begin{figure}
    \centering
    \includegraphics[width=9cm]{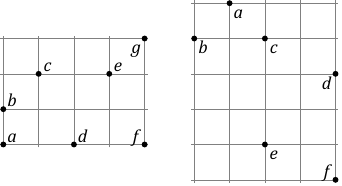}
    \caption{\label{fig_rigid}Left: The set $Q_7$, which satisfies $C$ but not $M$. Right: The set $Q_6$, which satisfies $P_4$ but not $C$.}
    
\end{figure}

\begin{lemma}
    The (non-all-even) set $Q_7=\{a,b,c,d,e,f,g\}$ where
    \begin{equation*}
        a=(0,0),\quad b=(0,1),\quad c=(1,2),\quad d=(2,0),\quad e=(3,2),\quad f=(4,0),\quad g=(4,3)
    \end{equation*}
    satisfies $C$ but not $M$.
\end{lemma}

\begin{proof}
    A contagious triangle in $Q_7$ is $\{a,b,c\}$. Since $ac\parallel de$, we merge the classes $\{d\}$ and $\{e\}$ into $\{d,e\}$. Since $bc\parallel eg$, we merge this latter class with $\{g\}$ into $\{d,e,g\}$. Since $ab\parallel fg$, we merge this latter class with $\{f\}$ into $\{d,e,f,g\}$. Finally $df\parallel ce$ allows us to merge this latter class with $\{a,b,c\}$.

    Our computer program verifies that $Q_7$ has no monotonically contagious triangle.
\end{proof}

\begin{lemma}
    The (non-all-even) set $Q_6=\{a,b,c,d,e,f\}$ where
    \begin{equation*}
    a=(1,5),\quad b=(0,4),\quad c=(2,4),\quad d=(4,3),\quad e=(2,1),\quad f=(4,0)
    \end{equation*}
    satisfies $P_4$ but not $P_3$.
\end{lemma}

\begin{proof}
    On the one hand, the initial pair of slopes $\{-1,1\}$ is $4$-contagious. Indeed, we can proceed as follows:
    \begin{itemize}
        \item The height-rigid subset $\{a,b,c,f\}$ (which contains slopes $-1,1$) infects slopes $-2,-5/3,0$.
        \item The height-rigid subset $\{c,d,e,f\}$ (which is height-rigid and contains slopes $-2,1$) infects slopes $-1/2,\infty$.
        \item Finally, subsets $\{a,c,d\}$, $\{b,c,e\}$, $\{a,c,e\}$, $\{b,c,d\}$ infect the remaining slopes $-2/3$, $-3/2$, $-4$, $-1/4$.
    \end{itemize}

    On the other hand, no triangle is contagious. Indeed, the following partitions of $Q_6$ are all terminal, in the sense that the triangle-contagion process cannot make any progress on them. 
    \begin{multline*}
        abcf\mid de, \quad abde\mid c\mid f,\quad acd\mid bef,\quad ace\mid bdf,\quad adf\mid b\mid ce,\\ aef\mid b\mid cd,\quad a\mid bcd\mid ef,\quad a\mid bce\mid df,\quad ab\mid cdef.
    \end{multline*}
    Furthermore, every possible initial triangle is contained in one of the classes of one of these partitions.
\end{proof}

We omit constructions that separate $P_k$ from $P_{k+1}$ for every $k\ge 4$.

\section{Constructions of all-even sets in $\R^2$}\label{sec:constructions}

Let $P_n=\set{(\cos(2\pi i/n),\sin(2\pi i/n))}{0\le i<n}$ denote the set of vertices of a regular $n$-gon. As Seoane-Sep\'ulveda showed, the set $P_{15}\setminus P_3$ is all-even.

In this section we present several other all-even sets, including some infinite families. The constructions presented here are of two types:
\begin{itemize}
    \item Subsets of a regular $n$-gon $P_n$. We call such sets \emph{cyclotomic}.
    \item Subsets of the integer grid $\mathbb Z^2$.
\end{itemize}

Regular polygons also arise in the study of planar point sets determining
few directions. Ungar~\cite{Unga82} proved that a non-collinear set of $n$
planar points determines at least $2\lfloor n/2\rfloor$ directions.
Among $n$-point sets with no three points collinear, Jamison~\cite{Jami86}
characterised those determining exactly $n$ directions as the affine
images of the vertices of a regular $n$-gon. Pilatte~\cite{Pila20} proved
that, for $n\geq 7$, a set of $n$ planar points with no three collinear
that determines exactly $n+1$ directions is affinely equivalent to the
vertex set of a regular $(n+1)$-gon with one vertex removed. These results
concern the total number of secant directions, whereas all-evenness
constrains the parity of the occupied-line count in every direction.

\subsection{Cyclotomic all-even sets}

Let $n$ be fixed, denote $[n]=\{0,\ldots,n-1\}$, and denote $p_i=(\cos(2\pi i/n), \sin(2\pi i/n))$ for $i\in[n]$. It is easy to verify that two pairs of points $p_ip_j$, $p_kp_\ell$ determine the same direction if and only if $i+j\equiv k+\ell\pmod n$. Hence, every direction determined by $P_n$ corresponds to an element of $[n]$. 

The following generalization of Seoane-Sep\'ulveda's example is also due to him and was presented by Santos at the same open problems session:

\begin{lemma}\label{lem_P_minus_P}
    For every $m\ge 1$ and $k\ge 0$, the set $A=P_{(4m+1)(4k+3)}\setminus P_{4k+3}$ is all-even.
\end{lemma}

\begin{proof}
    Let $n=(4m+1)(4k+3)$, and let
    \begin{equation*}
        T=\{(4m+1)i : 0\le i<4k+3\},
    \end{equation*}
    so $A$ consists of all the points $p_i = (\cos(2\pi i/n),\allowbreak \sin(2\pi i/n))$ for $i\in[n]\setminus T$.

    Consider a direction $\ell\in[n]$. Let $S_\ell$ be the set of all sets $\{i,j\}$ such that $i+j\equiv \ell \pmod n$. Since $n$ is odd, $S_\ell$ consists of $(n-1)/2$ pairs plus one singleton set $\{j\}$ with $2j\equiv\ell\pmod n$. Since $n\equiv 3 \pmod 4$, the set $S_\ell$ has even size. Now let us look at the set $T$ of omitted indices. We need to show that the number of sets in $S_\ell$ that are \emph{not} subsets of $T$ is even.
    
    If $4m+1\mid \ell$, then one element of $T$ is the index $j$ of the singleton set, and the remaining elements of $T$ appear together in $2k+1$ other pairs in $S_\ell$. Hence, the number of sets in $S_\ell$ that are subsets of $T$ is even, and so the number of sets in $S_\ell$ that are not subsets of $T$ is even as well, as required.

    Now suppose $4m+1\nmid\ell$. In this case, the $4k+3$ elements of $T$ appear in $4k+3$ different pairs in $S\ell$. Hence, no set in $S\ell$ is a subset of $T$, so the number of sets in $S_\ell$ that are not subsets of $T$ is even in this case as well.
\end{proof}

Let us represent a subset of the vertices of a regular $n$-gon by a binary string of length $n$, where a $0$ denotes a vertex not in the set, and a $1$ denotes a vertex in the set. For example, Seoane-Sep\'ulveda's original construction is given by the string $\mathtt{011110111101111}$.

\begin{figure}
    \centering
    \includegraphics{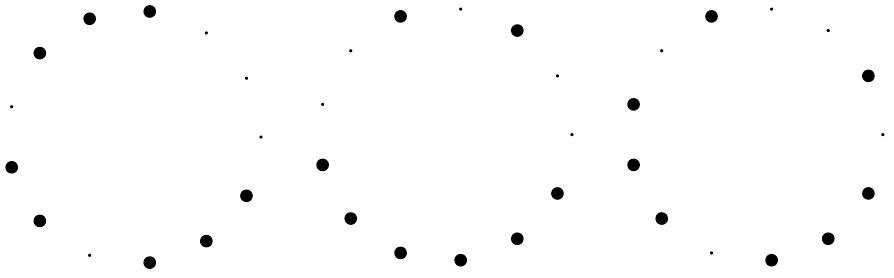}
    \caption{\label{fig_circ13}All-even sets that are subsets of $P_{13}$.}
\end{figure}

\begin{table}
    \begin{tabular}{ccccc}
    Binary string & $n$ & Size & M & Aux.\\\hline
    \texttt{0001110110111}&$13$&$8$&$\{3,4,7\}$&$\{3,4\}$\\
    \texttt{0010100111111}&$13$&$8$&$\{2,4,7\}$&$\{2,7\}$\\
    \texttt{0010111011101}&$13$&$8$&$\{2,4,6\}$&$\{2,4\}$\\
    \texttt{00001101111011}&$14$&$8$&$\{4,5,7\}$&$\{4,7\}$\\
    \texttt{0101011111011111}&$16$&$12$&$\{1,3,6\}$&$\{1,6\}$\\
    \texttt{000001001010010000011101010111}&$30$&$12$&$\{5,13,20\}$&$\{5,28\}$\\
    \texttt{000100010001110100100010010111}&$30$&$12$&$\{3,7,18\}$&$\{3,22\}$\\
    \texttt{0000100100111011011110001010111}&$31$&$16$&$\{4,7,11\}$&$\{4,7\}$\\
    \texttt{000100101110001001011100010110111}&$33$&$16$&$\{3,6,10\}$&$\{3,8\}$\\
    \texttt{00000010101001011101011001000101111}&$35$&$16$&$\{6,8,13\}$&$\{6,15\}$\\
    \texttt{00000011000010010011100011101011111}&$35$&$16$&$\{6,7,15\}$&$\{6,12\}$\\
    \texttt{00010001010010011000100100110111111}&$35$&$16$&$\{3,7,12\}$&$\{3,7\}$\\
    \end{tabular}
    \par\vspace{1em}
    \caption{\label{table_cyc}Some cyclotomic all-even sets.}
\end{table}

Table \ref{table_cyc} presents some cyclotomic all-even sets, encoded as binary strings. For each set, the table presents a monotonically contagious triangle (column ``M''), which proves that the set is height-rigid. Column ``Aux.'' pertains to Lemma \ref{lem_monotone_repeat_binary} below.

Figure \ref{fig_circ13} shows the first three sets in the table.

These sets were found by computer search (see code in the ancillary files). They exhaust, up to symmetry, all cyclotomic all-even sets not of the form $P_n\setminus P_k$ up to $n=35$.

In addition, every cyclotomic all-even set generates an infinite family of such sets:

\begin{lemma}\label{lem_repeat_binary}
    Let $a\in\{0,1\}^m$ be a binary string representing an all-even subset $A\subset P_m$. Then, for every $k\ge 0$, the subset $B\subset P_{m(4k+1)}$ represented by the binary string $b=a^{4k+1}=\underbrace{aa\cdots a}_{4k+1\text{ times}}$, is all-even as well.
\end{lemma}

\begin{proof}
    Let $I_a=\{i\in[m]:a_i=1\}$ be the set of indices corresponding to $A$. Then the set of indices corresponding to $B$ is $I_b=\{i+jm:i\in I_a, j\in[4k+1]\}$. Let $n=m(4k+1)$. For every $i\in[n]$ we have $i\in I_b$ if and only if $i\bmod m\in I_a$.

    Given $\ell\in[m]$, let $S_\ell$ be the set of sets $\{i,j\}$ with $i,j\in[m]$ and $i+j\equiv\ell\pmod m$. By assumption, for every $\ell\in[m]$, the set $\{s\in S_\ell:s\cap I_a\neq\emptyset\}$ has even size.
    
    Given a set of indices $\tau=\{i,j\}$ with $i,j\in[n]$, let $\overline\tau=\{i\bmod m, j\bmod m\}$.

    Given $\ell\in[n]$, let $T_\ell$ be the set of sets $\{i,j\}$ with $i,j\in[n]$ and $i+j\equiv\ell\pmod n$. We need to prove that for every $\ell\in[n]$, the set $\{t\in T_\ell:t\cap I_b\neq\emptyset\}$ has even size. Hence, let $\ell\in[n]$ be given, and denote $\ell'=\ell\bmod m$.

    Suppose first that $m=2m'+1$ is odd. Then $n$ is odd as well. Thus, $|T_\ell|=(n+1)/2$, and $T_\ell$ contains $(n-1)/2$ pairs plus one singleton $\{j\}$. Specifically, the elements of $T_\ell$ are exactly $\tau_0,\ldots,\tau_{(n-1)/2}$ where each $\tau_i = \{(j-i)\bmod n,(j+i)\bmod n\}$. Denote $\sigma_i=\{(j-i)\bmod m,(j+i)\bmod m\}$ for $0\le i\le m'$. Then the elements of $S_{\ell'}$ are exactly $\sigma_0,\ldots,\sigma_{m'}$.

    Now look at the sequence $\overline{\tau_0},\overline{\tau_1},\ldots,\overline{\tau_{(n-1)/2}}$. This sequence equals
    \begin{align*}
        \sigma_0,\ldots,\sigma_{m'},\sigma_{m'},\ldots,\sigma_0,\ldots,\sigma_{m'}, \quad\cdots\quad \sigma_{m'}.
    \end{align*}
    Here, $\sigma_0$ appears $2k+1$ times, and each one of $\sigma_1,\ldots,\sigma_{m'}$ appears $4k+1$ times. These multiplicities are all odd, so as far as parity is concerned, it is the same as if each one of $\sigma_0, \ldots,\sigma_{m'}$ appeared once. By assumption, an even number of them intersect $I_a$. Hence, an even number of $\tau_0,\ldots,\tau_{(n-1)/2}$ intersect $I_b$, as desired.

    Now suppose $m=2m'$ is even. Here there are two sub-cases, depending on whether $\ell$ is even or odd. If $\ell$ is even, then $T_\ell$ contains $n/2-1$ pairs plus two antipodal singletons. Let $\{j\}$ be one of the singletons. Then the elements of $T_\ell$ are exactly $\tau_0,\ldots,\tau_{n/2}$ where $\tau_i=\{(j-i)\bmod n,(j+i)\bmod n\}$ for each $i$. Denote $\sigma_i=\{(j-i)\bmod m,(j+i)\bmod m\}$ for $0\le i\le m'$. As before, the elements of $S_{\ell'}$ are exactly $\sigma_0,\ldots,\sigma_{m'}$.
    In this case, the sequence $\overline{\tau_0},\ldots,\overline{\tau_{n/2}}$ equals
    \begin{align*}
        \sigma_0,\ldots,\sigma_{m'},\ldots,\sigma_0,\ldots,\sigma_{m'},\quad\cdots\quad\sigma_{m'}.
    \end{align*}
    Here, $\sigma_0$ and $\sigma_{m'}$ appear $2k+1$ times each, and each one of $\sigma_1,\ldots,\sigma_{m'-1}$ appears $4k+1$ times. These multiplicities are all odd, so we argue as before.

    Finally, if $\ell$ is odd, then $T_\ell$ contains $n/2$ pairs and no singletons. Let $\{j,j+1\}$ be one of the pairs (there are two pairs of this form). Then the elements of $T_\ell$ are exactly $\tau_0,\ldots,\tau_{n/2-1}$ where $\tau_i=\{(j-i)\bmod n,(j+1+i)\bmod n\}$ for each $i$. Denote $\sigma_i=\{(j-i)\bmod m,(j+1+i)\bmod m\}$ for $0\le i< m'$. As before, the elements of $S_{\ell'}$ are exactly $\sigma_0,\ldots,\sigma_{m'-1}$. In this case, the sequence $\overline{\tau_0},\ldots,\overline{\tau_{n/2-1}}$ equals
    \begin{equation*}
        \sigma_0,\ldots,\sigma_{m'-1},\sigma_{m'-1},\ldots,\sigma_0,\sigma_0,\ldots,\sigma_{m'-1},\quad\cdots\quad\sigma_{m'-1}.
    \end{equation*}
    In this case each $\sigma_i$ appears $4k+1$ times. These multiplicities are all odd, so we proceed as before.
\end{proof}

\subsection{Sets with integer coordinates}

The following sets are all-even (see Figure \ref{fig_grid_AE_sets}), as our computer program (see ancillary files) verifies:

\begin{figure}
    \centering
    \begin{tabular}{cc}
        \makecell{\includegraphics[width=7cm]{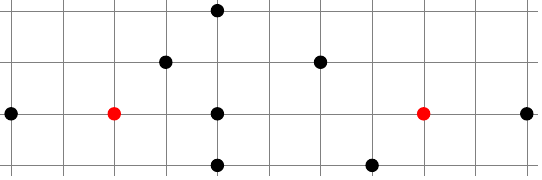}\\ \\\includegraphics[width=4cm]{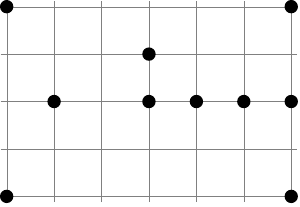}}&\makecell{\includegraphics[width=6cm]{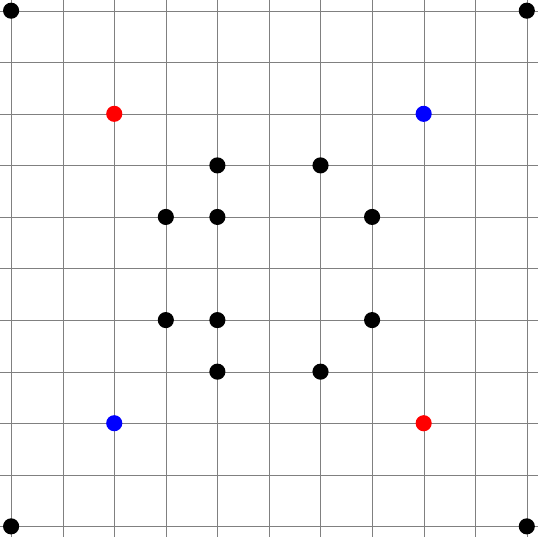}}
    \end{tabular}
    \caption{\label{fig_grid_AE_sets}Some integer-coordinate all-even sets. Top left: Sets $S_8$ (black points only) and $S_{10}$ (black and red points). Bottom left: Set $S_{10}'$. Right: Sets $S_{14}$ (black points), $S_{16}$ (black and blue points), and $S_{18}$ (black, blue, and red points).}
\end{figure}

\begin{align*}
    S_8 &=\{(0, -1), (3, -1), (-4, 0), (0, 0), (6, 0), (-1, 1), (2, 1), (0, 2)\},\\
    S_{10} &=S_8\cup\{(-2,0),(4,0)\},\\
    S_{10}' &=\{ (1,1),(7,1),(2,3),(4,3),(5,3),(6,3),(7,3),(4,4),(1,5),(7,5) \},\\
    S_{14} &=\{(-5, -5), (5, -5), (-1, -2), (1, -2), (-2, -1), (-1, -1), (2, -1),(-2, 1),\\&\qquad (-1, 1), (2, 1), (-1, 2), (1, 2), (-5, 5), (5, 5)\},\\
    S_{16} &=S_{14}\cup\{(-3,-3),(3,3)\},\\
    S_{18}&=S_{16}\cup\{(-3,3),(3,-3)\}.
\end{align*}

\begin{figure}
    \centering
    \includegraphics{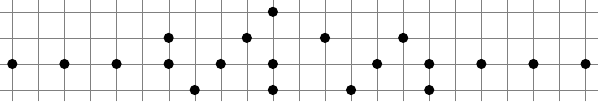}
    \caption{\label{fig_a1}The all-even set $A_k$ (here $k=1$).}
\end{figure}

Furthermore, the following set $A_k$, of size $10k+10$, is all-even for every $k\ge 1$ (see Figure \ref{fig_a1}):
\begin{align*}
    A_k&=\{(3i-1,-1):-k\le i\le k+1\}\\
    &{}\cup\{(2i-1,0):-3k-2\le i\le 3k+3,i\neq 1\}\\
    &{}\cup\{(3i+1,1):-k-1\le i\le k\}\\
    &{}\cup\{(-1,2)\}.
\end{align*}

\begin{lemma}\label{lem_Ak}
    The set $A_{k}$ is all-even for every $k\ge 1$.
\end{lemma}

\begin{proof}
    The set $A_k$ is covered by four horizontal lines, so we only need to address the non-horizontal directions.

    Consider the set $B=B_{-1}\cup B_0\cup B_{1}$ where:
    \begin{align*}
        B_{-1}&=\{(3i-1,-1):-k\le i\le k+1\},\\
        B_0&=\{(2i-1,0):-3k-2\le i\le 3k+3\},\\
        B_1&=\{(3i+1,1):-k-1\le i\le k\}.
    \end{align*}
    The set $B$ is centrally-symmetric around the origin. Hence, for every direction $U$ determined by $B$, the number of lines with direction $U$ that cover $B$ is odd if and only if one of these lines passes through the origin. This happens exactly for the directions of lines that pass through a point of $B_{-1}$ and the opposite point of $B_1$. The set of inverse-slopes of these directions is
    \begin{equation*}
        H = \{-3k-2, -3k+1, -3k+4,\ldots,3k+1\}.
    \end{equation*}
    Now, $A_k$ is obtained from $B$ by removing point $q=(1,0)$ and adding point $p=(-1,2)$.

    Let $J$ be the set of inverse-slopes of non-horizontal directions that lose no line by the removal of $q$. These are exactly the directions determined by $qr$ for some $r\in B_{-1}\cup B_1$. Hence,
    \begin{align*}
        J &= \{-3k-1,-3k+2,\ldots,3k+2\}\cup\{-3k-3,-3k,\ldots,3k\}\\
        &= I\setminus H,
    \end{align*}
    where
    \begin{equation*}
        I=\{-3k-3,-3k-2,-3k-1,\ldots,3k+2\}.
    \end{equation*}

    Now let us see which non-horizontal directions gain no line by the addition of $p$. These are exactly the directions determined by $pr$ for some $r\in B$. It is easy to check that the case $r\in B_0$ covers the cases $r\in B_{-1}$, $r\in B_1$ as well. Hence, the inverse-slopes of the directions that gain no line is exactly $I$.

    In summary, let $t$ be the inverse-slope of a non-horizontal direction.
    \begin{itemize}
        \item If $t\in H$, then the number of lines covering $B$ was odd, and the removal of $q$ and the addition of $p$ caused the loss of one line.
        \item If $t\in I\setminus H$, then the number of lines covering $B$ was even, and the removal of $q$ and the addition of $p$ caused no changes.
        \item If $t\notin I$, then the number of lines covering $B$ was even, and the removal of $q$ and the addition of $p$ caused the loss of one line and the gain of one line.
    \end{itemize}
    Hence, for every direction, the number of lines covering $A_k$ is even.
\end{proof}

\subsection{Contagion properties of these sets}\label{subsec:constructions_contagion_properties}
All the all-even sets we have presented in this section satisfy the strongest contagion property $M$ (existence of a monotonically contagious triangle). Our computer program (see ancillary files) verifies that each of the sets, except for the infinite families given by Lemmas \ref{lem_P_minus_P}, \ref{lem_repeat_binary}, and \ref{lem_Ak}, has a monotonically contagious triangle. We now prove that the infinite families given by Lemmas \ref{lem_P_minus_P} and \ref{lem_Ak}, as well as those given by applying Lemma \ref{lem_repeat_binary} to the binary strings of Table \ref{table_cyc}, satisfy property $M$ as well.

\begin{lemma}
Every cyclotomic set of the form $A=P_{km}/P_k$, $m\ge 5$, has a monotonically contagious triangle.
\end{lemma}

\begin{proof}
    Let $p_i=(\cos(2\pi i/n),\sin(2\pi i/n))$, and note that $\{p_1,p_2,p_3\}\in A$ and that $A$ is not missing any consecutive pair $p_i,p_{i+1}$. We claim that, starting with the initial triangle $\{p_1,p_2,p_3\}$, we can infect all subsequent points of $A$ in sequential order. Indeed, suppose we want to infect $p_i$, $i\ge 4$, and that we have already infected all previous points. If $p_{i-1}\in A$, then we use $p_1p_i\parallel p_2 p_{i-1}$. And if $p_{i-1}\notin A$, then $p_{i-2}\in A$, so we use $p_1 p_i\parallel p_3p_{i-2}$.
\end{proof}

Our next task is to prove that for every binary string $b$ listed in Table \ref{table_cyc} and every $k$, the cyclotomic set represented by $b^{4k+1}$ has a monotonically contagious triangle.

For this, we define an \emph{auxiliary contagion rule} for cyclotomic sets $A\subseteq P_n$ as follows: We start by taking an initial pair $U=\{p_i,p_j\}$ for some $p_i,p_j\in A$. Whenever $p_z\in A\setminus U$ and there exist elements $p_w,p_x,p_y\in U$ with $w+x\equiv y+z\pmod n$, where $w,x$ are \emph{not necessarily distinct}, we let $U\leftarrow U\cup\{p_z\}$. If we reach $U=A$ then we say that the initial pair $\{p_i,p_j\}$ is \emph{auxiliary-contagious}.

There are two differences between this rule and the normal monotonic contagion rule: Here we start with a pair instead of a triple, and here we allow $w=x$ whereas in the normal rule this is disallowed.

\begin{lemma}\label{lem_monotone_repeat_binary}
 Let $p_0,\ldots,p_{n-1}$ be the points of $P_n$ listed in the standard order. Suppose $\{p_i,p_j\}$ is an auxiliary-contagious pair in the cyclotomic set $A\subseteq P_n$. Let $b\in\{0,1\}^n$ be the binary string representing $A$, and let $k\ge 2$. Then $\{q_i,q_j,q_{i+n}\}$ is a monotonically contagious triangle in the cyclotomic set $B\subseteq P_{nk}$ corresponding to $b^k$, where $q_0,\ldots,q_{nk-1}$ are the points of $P_{nk}$ listed in the standard order.
\end{lemma}

\begin{proof}
    Start the contagion process of $B$ by infecting $q_{j+n}, q_{j+2n}, \ldots, q_{j+(k-1)n}$ via $q_{i+n}q_{j+un}\parallel q_iq_{j+(u+1)n}$ for $u=0,1,\ldots,k-2$ (indices are modulo $kn$). Using a similar approach, infect $q_{i+2n},q_{i+3n},\ldots,q_{i+(k-1)n}$. In general, we can ensure that whenever some $q_u$ is infected, all $q_{u'}$ with $u'\equiv u \pmod n$ are infected as well.

    We emulate the auxiliary contagion process on $A$. Whenever the auxiliary process uses $p_w,p_x,p_y\in A$ to infect $p_z\in A$ via $w+x\equiv y+z \pmod n$ (in which we could have $w=x$), we take some indices $w', x', y'\in[nk]$ such that $w'\equiv w \pmod n$, $x'\equiv x \pmod n$, $y'\equiv y \pmod n$, $z'\equiv z \pmod n$, and such that $w'\neq x'$ (which is possible since $k\ge 2$) to infect $q_{z'}$ via $q_{w'}q_{x'}\parallel q_{y'}q_{z'}$. Then we infect all $q_{z''}$ with $z''\equiv z' \pmod n$, and we continue with the next step of the process on $A$.
\end{proof}

Since every binary string in Table \ref{table_cyc} contains an auxiliary-contagious pair, as shown in the last column of the table, every cyclotomic set generated from these strings via Lemma \ref{lem_repeat_binary} is monotonically contagious, as claimed.

\begin{lemma}
    For every $k\ge 1$, the set $A_k$ of Lemma \ref{lem_Ak} has a monotonically contagious triangle.
\end{lemma}

\begin{proof}
    The triangle $\{a,b,c\}$ with $a=(-1,2)$, $b=(-2,1)$, $c=(1,1)$ is monotonically contagious. Indeed, the direction $ab$ allows us to infect points $(-3,0)$ and $(-4,-1)$. Then the horizontal direction $bc$ allows us to infect all the remaining points.
\end{proof}

\section{Singly-determined directions and their significance}\label{sec_singly}

Let $S\subset\R^d$ be a finite set. We say that $S$ \emph{singly-determines} direction $\ell\in\mathbb P(\R^d)$ if there exists a unique pair of points $p,q\in S$ that determine direction $\ell$.

\begin{lemma}\label{lem_singly}
    If $S\subset \R^d$ singly-determines some direction $\ell$, then $S$ has an odd line direction.
\end{lemma}

\begin{proof}
    We have already shown that if $|S|$ is odd then $S$ has an odd line direction. If $|S|$ is even, then the singly-determined direction $\ell$ is an odd line direction.
\end{proof}

In particular, if $S\subset \R^2$ singly-determines some direction, then $S$ is not all-even, and if $S\subset \R^3$ singly-determines some direction, then $S$ is not line-all-even (and hence not all-even either).

\begin{lemma}\label{lem_2lines}
    Let $S\subset \R^2$ be all-even. Then $n_S(\ell)\ge 4$ for all $\ell\in\mathbb P(\R^2)$.
\end{lemma}

\begin{proof}
    Suppose for a contradiction that there exists a direction $\ell\in\mathbb P(\R^2)$ with $n_S(\ell)=2$. Let $L_1, L_2$ be the two lines with direction $\ell$ that cover $S$. Let $u\in\Sph^1$ be one of the sub-directions parallel to $\ell$. Let $p$ be the first point of $S\cap L_1$ according to sub-direction $u$, and let $q$ be the last point of $S\cap L_2$ according to sub-direction $u$. Then it is easy to check that the direction determined by $pq$ is a singly-determined direction. Hence, the claim follows by Lemma \ref{lem_singly}.
\end{proof}

\begin{proposition}
    Let $S\subset\mathbb R^3$ be all-even. Then $m_S(\ell)\ge 4$ for all $\ell\in\mathbb P(\R^3)$.
\end{proposition}

\begin{proof}
    Suppose for a contradiction that there exists a direction $\ell\in\mathbb P(\R^3)$ with $m_S(\ell)=2$. Let $\pi_1$, $\pi_2$ be the two planes perpendicular to $\ell$ that cover $S$. Let $u$ be a line direction perpendicular to $\ell$. Suppose for simplicity that $\pi_1$, $\pi_2$ are horizontal (parallel to the $xy$-plane) and that $u$ is parallel to the $x$-axis. Let $U_1$ be the lowest (in $y$-coordinate) line with direction $u$ that contains a point of $S\cap \pi_1$, and let $U_2$ be the highest (in $y$-coordinate) line with direction $u$ that contains a point of $S\cap \pi_2$. Let $p$ be the leftmost (smallest $x$-coordinate) point of $S\cap U_1$, and let $q$ be the rightmost (largest $x$-coordinate) point of $S\cap U_2$. It can be seen that the direction $pq$ is singly determined: Indeed, let $v$ be the vector $q-p$. Any other vector $v'=q'-p'$, with $p'\in S\cap\pi_1$ and $q'\in S\cap\pi_2$, will have the same $z$-coordinate as $v$, meaning $v_z=v'_z$. If $p'\notin U_1$ or $q'\notin U_2$, then $v'_y<v_y$. And if $p\in U_1$ and $q\in U_2$ then $v'_y=v_y$ but $v'_x<v_x$. In any case, $p'q'$ is not parallel to $pq$. Hence, the claim follows from Lemma \ref{lem_singly}.
\end{proof}

\section{A six-line theorem}\label{sec_6lines}

In this section we prove the following:

\begin{theorem}\label{thm_6lines}
Let $A\subset\R^2$ be all-even, and suppose there exists a direction $\ell\in\PP(\R^2)$ for which $n_A(\ell)\le 6$. Then $A$ is height-rigid.
\end{theorem}

Hence, if there exists any all-even set $A\subset\R^2$ that is not height-rigid, it must satisfy $n_A(\ell)\ge 8$ for all directions $\ell$. Therefore, we must have $|A|\ge 10$: If $|A|\le8$ then a secant direction $\ell$ would yield $n_A(\ell)<8$, and thus $n_A(\ell)\le 6$.

The rest of this section proves Theorem \ref{thm_6lines}.

Let $\ell$ be the assumed direction with $k=n_A(\ell)\le 6$, and let $L_1, \ldots, L_k$ be the $k$ lines with direction $\ell$ that cover $A$. By Lemma \ref{lem_2lines} we must have $k\ge 4$, so $k\in\{4,6\}$.

Suppose for a contradiction that $A$ has an admissible lift $z$ such that the lifted set $\hat A\subset\R^3$ is not coplanar. Then $\hat A$ is covered by $k$ parallel but not coplanar lines $\hat L_1, \ldots, \hat L_k$. Let $\Pi\subset\R^3$ be a plane perpendicular to these lines, and let the points $p_1,\ldots,p_k\in\Pi$ be the orthogonal projections of $\hat L_1,\ldots,\hat L_k$ into $\Pi$, respectively. Then $p_1,\ldots,p_k$ are not collinear.

\begin{figure}
    \centering\includegraphics[width=11cm]{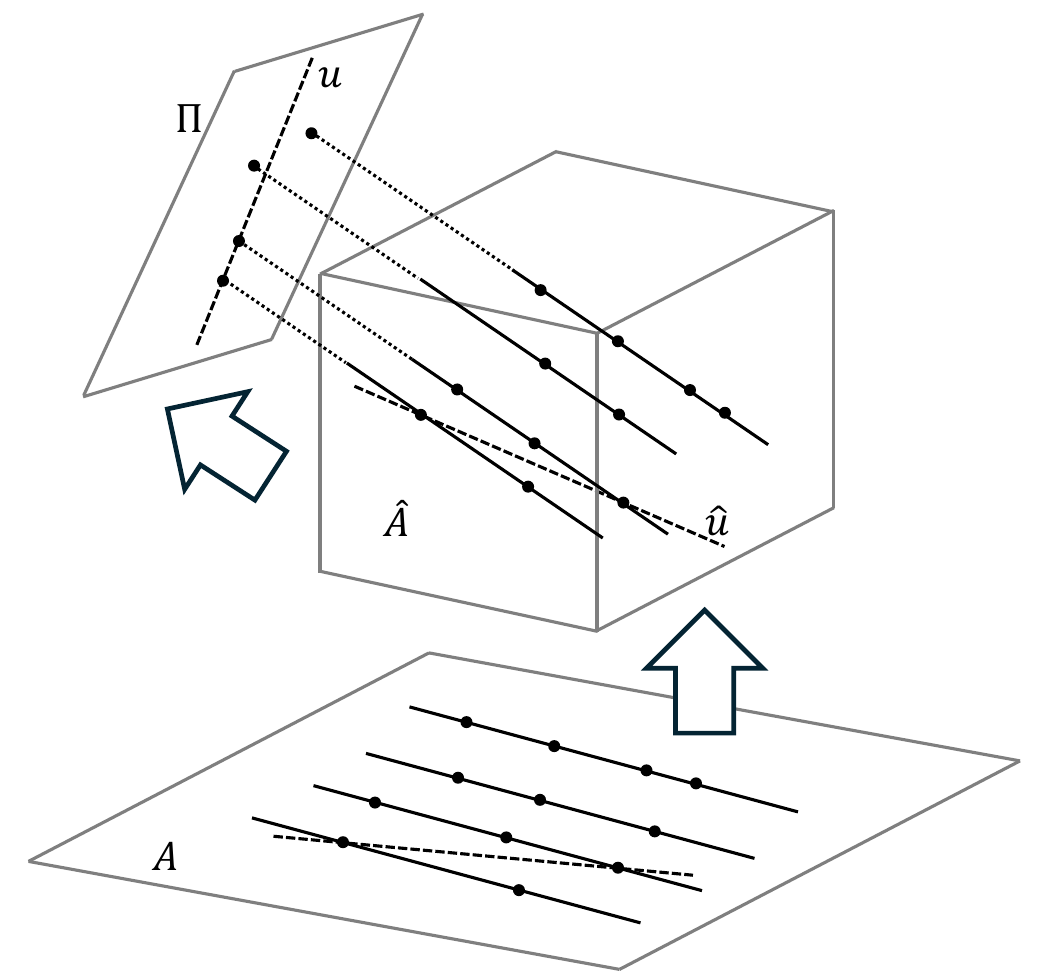}
    \caption{\label{fig_lift_and_project}A singly-determined direction in the plane $\Pi$ yields a singly-determined direction in the set $\hat A$, which in turn yields a singly-determined direction in the original set $A$.}
\end{figure}

Suppose first that some direction $u=\spn(p_ip_j)$ within $\Pi$ is singly determined in $\{p_1,\ldots,p_k\}$. Going back to $\hat A$, let $\hat q$ be the first point of $\hat A\cap\hat L_i$, and let $\hat r$ be the last point of $\hat A\cap\hat L_j$, where ``first'' and ``last'' are determined according to some arbitrary sub-direction of the direction of the lines $ L_i$. It is easy to see that the direction $\hat u$ determined by $\hat q\hat r$ is singly determined by $\hat A$. Since the lift $z$ is admissible, the source points $q,r\in A$ of $\hat q$, $\hat r$ determine a singly-determined direction in $A$. But this contradicts Lemma \ref{lem_singly}. See Figure \ref{fig_lift_and_project}.

Hence, suppose first that $k=4$. As a simple case analysis shows, given four non-collinear points in the plane, there is always a direction that is singly determined by them. Hence, the assumption $k=4$ readily yields a contradiction.

Now suppose $k=6$. A computer-assisted enumeration, outlined in Appendix \ref{app_180} below, shows that the only planar six-point sets that are non-collinear and do not have a singly-determined direction are the following, up to affine transformation:
\begin{itemize}
\item The set $G=\{(\cos \frac{2\pi i}{7},\sin \frac{2\pi i}{7}) : 0< i <7\}$ of six out of seven vertices of a regular heptagon.
\item The set $H=\{(\cos \frac{\pi i}{3},\sin \frac{\pi i}{3}) : 0\le i< 6\}$ of vertices of a regular hexagon.
\end{itemize}

We now analyze each of these cases separately.

\subsection{The regular heptagon without a vertex}

Let us work in a coordinate system $x'$, $y'$, $z'$, where the plane $\Pi$ is the $x'y'$-plane, and the lines $\hat L_1,\ldots,\hat L_6$ are parallel to the $z'$-axis.

\begin{figure}
    \centering
    \includegraphics[width=14cm]{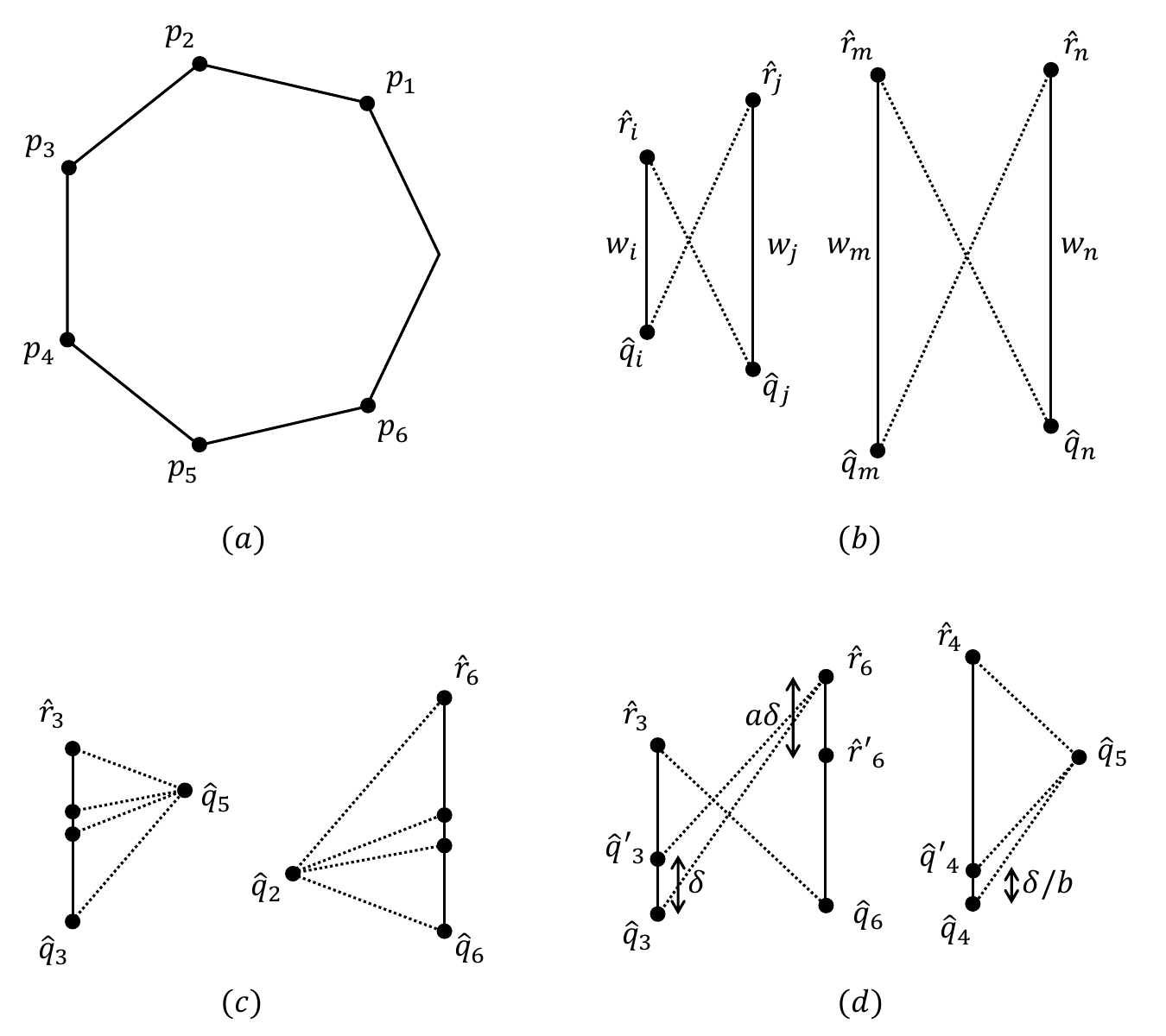}
    \caption{\label{figs_heptagon}Figures for the case of the regular heptagon without a vertex.}
\end{figure}

For concreteness, let $G=\{p_1,\ldots,p_6\}$ where $p_i=(\cos \frac{2\pi i}{7}, \sin \frac{2\pi i}{7},0)$ for $1\le i\le 6$. See Figure \ref{figs_heptagon}$(a)$. Let $a=\|p_2p_6\|/\|p_3p_5\| =(\sin\frac{3\pi}{7})/(\sin\frac{2\pi}{7})\approx 1.246$, and let $b=\|p_3p_6\|/\|p_4p_5\| =(\sin\frac{3\pi}{7})/(\sin\frac{\pi}{7})\approx 2.246$. We have $b=a+1$, since
\begin{multline*}
    b-a=\frac{\sin\frac{3\pi}{7}}{\sin\frac{\pi}{7}} - \frac{\sin\frac{3\pi}{7}}{\sin\frac{2\pi}{7}}=\frac{\sin\frac{3\pi}{7}\sin\frac{2\pi}{7}-\sin\frac{3\pi}{7}\sin\frac{\pi}{7}}{\sin\frac{\pi}{7}\sin\frac{2\pi}{7}}\\
    =\frac{\cos\frac{\pi}{7}-\cos\frac{5\pi}{7}-\cos\frac{2\pi}{7}+\cos\frac{4\pi}{7}}{\cos\frac{\pi}{7}-\cos\frac{3\pi}{7}}=1;
\end{multline*}
where we have used the identities $\sin \alpha\sin\beta=\frac{1}{2}(\cos(\alpha-\beta)-\cos(\alpha+\beta))$ and $\cos\alpha=-\cos(\pi-\alpha)$.

Call a direction $u$ in $\Pi$ \emph{twice-determined} in $G$ if $u$ is determined by $p_ip_j$ and by $p_mp_n$, where $i,j,m,n$ are four distinct indices, and $u$ is not determined by any other pair of points of $G$.

For each $1\le i\le 6$, let $\hat q_i$, $\hat r_i$ be the lowest and highest points in $\hat A\cap \hat L_i$, respectively (with respect to $z'$-coordinate); and let $w_i=\|\hat q_i\hat r_i\|$. The distances $w_i$ satisfy the following property:

\begin{lemma}\label{lem_wi}
Let $u$ be a direction twice-determined in $G$, determined by $p_ip_j$ and by $p_mp_n$. Then $w_i+w_j=c(w_m+w_n)$ where $c=\|p_ip_j\|/\|p_mp_n\|$.
\end{lemma}

\begin{proof}
    Consider the directions determined by $\hat q_i\hat r_j$ and $\hat q_m\hat r_n$. Clearly, the direction $\hat q_i\hat r_j$ is singly-determined in $\hat A\cap (\hat L_i\cup \hat L_j)$. Similarly, the direction $\hat q_m\hat r_n$ is singly-determined in $\hat A\cap (\hat L_m\cup\hat L_n)$. Hence, these two directions must be equal: If, say $\hat q_i\hat r_j$ were steeper in the $z'$-direction than $\hat q_m\hat r_n$, then $\hat q_i\hat r_j$ would be singly-determined in $\hat A$. Conversely, if $\hat q_m\hat r_n$ were steeper than $\hat q_i\hat r_j$, then $\hat q_m\hat r_n$ would be singly-determined in $\hat A$.

    Similarly, the directions determined by $\hat r_i\hat q_j$ and $\hat r_m\hat q_n$ must be equal.

    In other words, $T_1=\hat q_i\hat r_i\hat q_j\hat r_j$ and $T_2=\hat q_m\hat r_m\hat q_n\hat r_n$ are crossed trapezoids, where each side of $T_1$ is parallel to one side of $T_2$ and vice versa; and the ratio between the heights of the trapezoids is $c$. See Figure \ref{figs_heptagon}$(b)$. It follows that $w_i+w_j=c(w_m+w_n)$.
\end{proof}

Lemma \ref{lem_wi} can be applied to every direction determined by $G$ except for the direction determined by $p_1p_6\parallel p_2p_5\parallel p_3p_4$. We obtain the following linear system of equations on $w_1,\ldots,w_6$:
\begin{align*}
    w_2+w_6&=a(w_3+w_5),\\
    w_1+w_5&=a(w_2+w_4),\\
    w_3+w_6&=(a+1)(w_4+w_5),\\
    w_1+w_4&=(a+1)(w_2+w_3),\\
    w_4+w_6&=\frac{a+1}{a}(w_1+w_2),\\
    w_1+w_3&=\frac{a+1}{a}(w_5+w_6).
\end{align*}
These equations yield:
\begin{align}
    w_1&=w_6,\nonumber\\
    w_3&=w_4=w_1/a,\label{eq_w_sol}\\
    w_2&=w_5=0.\nonumber
\end{align}
In particular, each one of $\hat  L_2$, $\hat L_5$ must contain a single point of $\hat A$, namely $\hat q_2=\hat r_2$ and $\hat q_5=\hat r_5$.

Considering the twice-determined direction $p_3p_5\parallel p_2p_6$, we conclude that the triangles $\hat q_3\hat r_3\hat q_5$ and $\hat r_6\hat q_6\hat q_2$ are antiparallel and similar, under a scaling factor of $a$. Furthermore, the point set $\hat A\cap \hat L_6$ must be an inverted and scaled-up copy of the point set $\hat A\cap \hat L_3$, with a scaling factor of $a$, since otherwise there would be a singly-determined direction. See Figure \ref{figs_heptagon}$(c)$.

Similarly, by considering the twice-determined direction $p_2p_4\parallel p_1p_5$, we conclude that the triangles $\hat q_2\hat q_4\hat r_4$ and $\hat q_5\hat r_1\hat q_1$ are antiparallel and similar, under a scalar factor of $a$. Furthermore, the point set $\hat A\cap \hat  L_1$ must be an inverted and scaled-up copy of the point set $\hat A\cap\hat L_4$, with scaling factor $a$.

Now we claim that $\hat A\cap \hat L_3$ must contain a single point, so $w_3=0$. Indeed, assuming otherwise, let $\hat q'_3$ be the second-lowest point of $\hat A\cap\hat L_3$. Let $\delta=\|\hat q_3\hat q'_3\|$. We have already seen that the direction $\hat q_3\hat r_6$ is singly-determined in $\hat A\cap (\hat L_3\cup\hat L_6)$. The same must be true of the direction $\hat q'_3\hat r_6$, since $\hat r'_6$, the second-highest point of $\hat A\cap\hat  L_6$ is at distance $a\delta$ from $\hat r_6$, which is too far away to determine the same direction.

Hence, the direction $\hat q'_3\hat r_6$ must be determined a second time by some point in $\hat A\cap\hat  L_4$ and $\hat q_5$. This implies that  the second-lowest point $\hat q'_4$ of $\hat A\cap \hat  L_4$ must satisfy $\|\hat q_4\hat q'_4\|\le\delta/b$. See Figure \ref{figs_heptagon}$(d)$.

But now, the same argument, this time applied to the twice-determined direction $p_4p_1\parallel p_3p_2$, yields that the second point $\hat q'_3$ of $\hat A\cap \hat L_3$ satisfies $\|\hat q_3 \hat q'_3\|\le\|\hat q_4\hat q'_4\|/b$. Thus, $\delta\le\delta/b^2$. This contradiction proves that $\hat A\cap \hat  L_3$ contains a single point, so $w_3=0$.

Equations (\ref{eq_w_sol}) imply that $w_1=\cdots=w_6=0$.

Hence, $\hat A$ consists of just the six points $\hat q_1,\ldots,\hat q_6$. Finally, consider the directions $\hat q_3\hat q_4$, $\hat q_2\hat q_5$, $\hat q_1\hat q_6$. If these three are the same direction, then this is an odd direction in $\hat A$, contradicting the fact that $\hat A$ is line-all-even. Otherwise, one of these directions must be different from the other two, so it is singly-determined by $\hat A$.

\subsection{The regular hexagon}\label{subsec:reg_hex}
As before, let us work in a coordinate system $x'$, $y'$, $z'$, where the plane $\Pi$ is the $x'y'$-plane, and the lines $\hat L_1,\ldots,\hat L_6$ are parallel to the $z'$-axis.

\begin{figure}
    \centering
    \includegraphics[width=7cm]{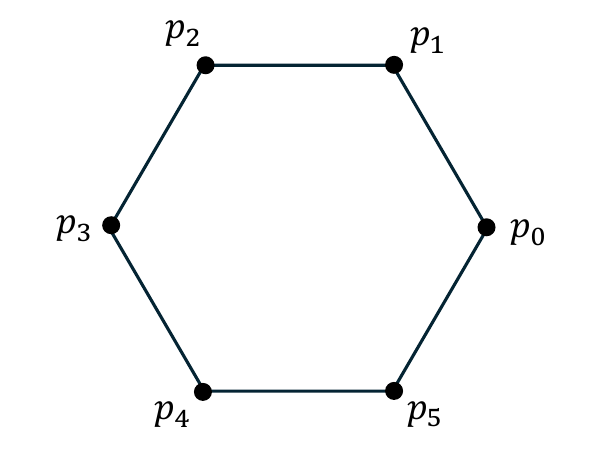}
    \caption{\label{fig_hexagon}The case of the regular hexagon.}
\end{figure}

For concreteness let $H=\{p_0,\ldots,p_5\}$ where $p_i=(\cos \frac{\pi i}{3}, \sin \frac{\pi i}{3},0)$ for $0\le i\le 5$.

For each $0\le i\le 5$, let $T_i\subset\R$ be the set of $z'$-coordinates of the points in $\hat A\cap\hat L_i$. We will now encode each $T_i$ as a real-exponent polynomial, and we will exploit properties of the ring $\mathbb F_2[\R]$ of real-exponent polynomials with $0/1$-coefficients modulo $2$. (A similar technique is used, for example, in \cite{rosenblatt-seymour}, though there the coefficients are in $\mathbb Z$ and not in $\mathbb F_2$.)

Note that this ring is actually an integral domain, meaning if $F_1, F_2\in \mathbb F_2[\R]$ are two polynomials that satisfy $F_1F_2=0$, then one of $F_1$, $F_2$ must be the zero polynomial. (Proof: Assuming a contradiction, consider the product of the highest term of each $F_i$. This term cannot cancel out.) Hence, if $F_1\neq 0$ and $F_1F_2=F_1F_3$, then $F_2=F_3$.

For each $0\le i\le 5$, define the real-exponent polynomial $F_i(X)=\sum_{z\in T_i} X^z$. Also define $F^*_i(X) = F_i(X^{-1})$, and for $c\in\R$, $c\neq 0$, define $D_c F_i(X) = F_i(X^c)$.

For example, if $T_i=\{1.5,2,3\}$, then $F_i(X) = X^{1.5}+X^2+X^3$, $F^*_i(X)=X^{-1.5}+X^{-2}+X^{-3}$, and $D_{1.1}F_i(X) = X^{1.65}+X^{2.2}+X^{3.3}$.

These polynomials satisfy the following property: Consider a planar sub-direction $u$ within the plane $\Pi$ determined by two points $p_ip_j$. Let $J_u=\{(i_1,j_1),\ldots,(i_k,j_k)\}$ be the set of all pairs of indices such that $p_{i_m}p_{j_m}$ determine sub-direction $u$. Now consider a spatial sub-direction $\hat u$ determined by two points $q\in\hat A\cap \hat  L_i$, $q'\in\hat A\cap \hat L_j$, $(i,j)\in J_u$. By which other pairs of points $r,r'\in\hat A$ is the same sub-direction $\hat u$ determined? One possibility is to have $r\in\hat L_i$ and $r'\in\hat L_j$. In this case, the $z$-coordinates of the points must have the same difference: $r'_z-r_z=q'_z-q_z$. Another possibility is to have $r\in\hat L_{i'}$, $r'\in\hat L_{j'}$ for some other pair $(i',j')\in J_u$. In this case, the $z$-coordinates must satisfy $r'_z-r_z=c(q'_z-q_z)$ where $c = \|p_{i'}p_{j'}\|/\|p_ip_j\|$. Since $\hat A$ is line-all-even and has an even number of points, and since no three points $p_i$ lie on a line, there must be an even number of pairs of points of $\hat A$ that determine sub-direction $\hat u$. This must be true for every spatial sub-direction $\hat u$ that is a lifting of $u$.

As a consequence, for every sub-direction $u$ within $\Pi$, the polynomials $F_i(X)$ must satisfy
\begin{equation*}
    \sum_{(i,j)\in J_u} D_{1/|p_ip_j|}(F^*_iF_j)=0,
\end{equation*}
where we work in $\mathbb F_2[\R]$, meaning the coefficients are modulo $2$.

In particular, we must have
\begin{align}
    F^*_1F_3&=F^*_0F_4,\label{eq_F_1}\\F^*_2F_4&=F^*_1F_5,\label{eq_F_2}\\
    F^*_2F_3 + D_{1/2}(F^*_1F_4)+F^*_0F_5&=0\label{eq_F_3}
\end{align}
Multiplying (\ref{eq_F_1}) by $F^*_2$, substituting (\ref{eq_F_2}), and canceling out $F^*_1$ (which is not the zero polynomial), we get
\begin{equation*}
    F^*_2F_3 = F^*_0F_5.
\end{equation*}
Substituting this latter equation into (\ref{eq_F_3}) we get $D_{1/2}(F^*_1F_4)=0$, which implies $F^*_1F_4=0$. Hence, $F^*_1=0$ or $F_4=0$, contradiction.

Together with Appendix \ref{app_180}, this concludes the proof of Theorem \ref{thm_6lines}.

\section{Concluding remarks and open problems}\label{sec:concluding}

By Theorem~\ref{thm:slc_equiv_AEisHR}, the strong conjecture (Conjecture \ref{conj:slc}) is equivalent
to the assertion that every planar all-even set is height-rigid.
The results above suggest several ways to approach this assertion:
contagion from small rigid subsets, analysis of the rigidity matrix,
and restrictions arising from configurations supported on few parallel
lines. We conclude with questions prompted by these approaches and by
the constructions of Section~\ref{sec:constructions}.

\subsection*{Contagion and height-rigidity}

The finite examples and the explicitly analysed infinite families of
Section~\ref{subsec:constructions_contagion_properties} admit monotonically
contagious triangles, and hence satisfy the strongest condition in the
hierarchy of Section~\ref{sec_contagion}. This suggests the following
strengthening of the strong conjecture.

\begin{conjecture}
Every planar all-even set $A$ has a monotonically contagious triangle.
\end{conjecture}

An affirmative answer would show that every planar all-even set admits
a particularly simple certificate of height-rigidity: starting from
three non-collinear points, the remaining points can be added one at a
time using parallelisms with previously determined secants.
By Theorem~\ref{thm:slc_equiv_AEisHR} and
Proposition~\ref{prop:AE_is_LAE}, this would also settle the strong
conjecture and answer Seoane-Sep\'ulveda's original question.

The hierarchy provides weaker possible sufficient conditions for the
strong conjecture. One may ask whether every planar all-even set
satisfies $C$, or, more generally, whether there is an absolute integer
$k\geq 3$ such that every such set satisfies $P_k$. The latter asks
whether height-rigidity can always be certified by propagation through
rigid subsets of uniformly bounded size. The separation examples in
Section~\ref{subsec:sep_contagion_properties} show that these
distinctions matter for general planar point sets. It remains open
whether any of these separations can occur among all-even sets.

\subsection*{The algebraic approach}

Section~\ref{sec:rigidity_matrix} expresses both all-evenness and
height-rigidity in terms of the rigidity matrix. For an all-even set
$A$ of size $n$, every secant-direction block $R_\ell(A)$ has odd rank,
while height-rigidity is equivalent to
\[
\operatorname{rank}R(A)=n-3.
\]
The difficulty is to pass from the parity of the individual directional
ranks to this global rank statement. The directional row spaces may
intersect, so their dimensions cannot simply be added. A successful
argument must therefore use the compatibility imposed by their common
geometric origin.

This suggests studying how the parallelism relations of a planar
configuration constrain the dependencies between its directional row
spaces. The reconstruction results of
Section~\ref{sec:rigidity_matrix} show that the full row space retains
the secant-direction blocks and the occupied-line partitions.
Thus the algebraic formulation preserves the data defining
all-evenness; the remaining task is to understand why those data
should force the space of admissible heights to consist only of affine
functions. A direct argument of this kind could establish the strong
conjecture without resolving the stronger contagion questions.

\subsection*{Constructions and contagion}

The cyclotomic and integral classes considered in
Section~\ref{sec:constructions} provide natural settings in which to
study the rigidity conjectures. Establishing height-rigidity for every
cyclotomic all-even set, or for every integral all-even set, would
already give substantial special cases of the strong conjecture.
In each class, one may also ask whether the stronger property $M$
always holds.

The repetition construction raises a more specific question.
Lemma~\ref{lem_repeat_binary} shows that repeating the binary word of
a cyclotomic all-even set $4k+1$ times preserves all-evenness.
Does this operation also preserve $M$ whenever the original set
satisfies $M$? Lemma~\ref{lem_monotone_repeat_binary} establishes this
conclusion under the additional hypothesis that the original set has
an auxiliary-contagious pair; in fact, that hypothesis ensures $M$
after any repetition with at least two copies. The remaining question
is whether the auxiliary-pair hypothesis can be replaced by $M$ for
the all-even repetitions considered here.

\subsection*{Beyond six parallel lines}

Theorem~\ref{thm_6lines} establishes height-rigidity when an all-even set
is supported on at most six parallel lines, with no bound on its
cardinality. Consequently, every planar all-even set $A$ that is not
height-rigid must satisfy
\[
n_A(\ell)\geq 8
\qquad\text{for every }\ell\in\mathbb{P}(\mathbb{R}^2).
\]

\begin{problem}
Is every planar all-even set $A$ satisfying $n_A(\ell)\leq 8$ for some
direction $\ell$ height-rigid? More generally, extend
Theorem~\ref{thm_6lines} to larger even bounds on the number of occupied
parallel lines.
\end{problem}

The proof of Theorem~\ref{thm_6lines} combines information about the
projected configuration with restrictions on the point sets carried
by the parallel lines of an admissible lift. A singly-determined
direction in the projected configuration is immediately excluded.
For six non-collinear projected points, Appendix~\ref{app_180}
reduces the remaining possibilities to two affine types, which are
then treated using extremal points and polynomial identities.

At eight projected points, a new feature appears: the configurations
with no singly-determined direction already include planar all-even
sets, as the eight-point examples of Section~\ref{sec:constructions}
demonstrate. The aim is to identify conditions on the point sets
carried by the parallel lines that force the lifted set to be
coplanar, preferably without a separate analysis of every projected
configuration.

The encoding by real-exponent polynomials over $\mathbb{F}_2$ in
Section~\ref{subsec:reg_hex} offers one possible ingredient.
It translates parity constraints into identities in an integral
domain, allowing cancellation. The pair-count identities used there
rely on the absence of collinear triples in the projected
configuration. Extending the method to configurations with collinear
triples therefore requires an encoding that accounts for the
contribution of lines containing more than two points. Such an
extension may help treat larger projected configurations.
An extension of Theorem~\ref{thm_6lines} to every even bound would
settle the strong conjecture, since for each all-even set $A$ one
may take the bound to be $|A|$.

\subsection*{Acknowledgements}
We thank Francisco Santos for personal conversations.

We used an LLM for computational exploration, algebraic checks, and editorial assistance. Some of the mathematical ideas and research directions emerged through extended discussions with the model. The authors take full responsibility for the mathematical content and its presentation.

\bibliographystyle{alpha}
\bibliography{aes}

\appendix

\section{Classification of six-point sets with no singly-determined direction}\label{app_180}

In this section we give the computer-assisted proof that the sets $G$ and $H$ of Section \ref{sec_6lines} (the regular heptagon without a vertex and the regular hexagon) are, up to affine transformation, the only planar, non-collinear six-point sets without a singly-determined direction. In particular, since neither $G$ nor $H$ are all-even, this implies that there exist no planar all-even sets of size $6$.

We first use the Goodman--Pollack allowable-sequence framework (see e.g.~\cite{goodman-pollack,padrol-philippe,goodman-pollack-chapter}) to obtain a computer-generated list of candidates, and then we show that the only candidates that are geometrically realizable are the above-mentioned sets $G$ and $H$.

Let $A\subset\R^2$ be an $n$ point set, and suppose $A$ is oriented in such a way that no two points have the same $x$-coordinate. Number the points of $A$ as $0, \ldots, n-1$ by increasing $x$-coordinate. Now let us see what happens to the $x$-order of the points as $A$ turns counterclockwise $180$ degrees. For each $0\le\alpha\le\pi$, let $A(\alpha)$ be a copy of $A$ rotated counterclockwise by $\alpha$ radians. If $\alpha$ is generic (no two points of $A(\alpha)$ have the same $x$-coordinate), then let $\sigma(\alpha)\in S_n$ be the permutation of the points of $A(\alpha)$ listed by increasing $x$-coordinate. Hence, as $\alpha$ increases from $0$ to $\pi$, $\sigma(\alpha)$ undergoes a finite number of discrete changes, with $\sigma(0)=(0,\ldots, n-1)$ and $\sigma(\pi)=(n-1,\ldots, 0)$.

The following properties are easily established:
\begin{itemize}
    \item Every combinatorial change at a critical angle $\alpha$ consists of a simultaneous reversal of one or more adjacent blocks in the permutation. These blocks are determined by the distribution of points along the lines that become vertical at angle $\alpha$.
    \item Whenever a block is reversed, the elements were in increasing order before the reversal, and are in decreasing order after the reversal. This is because every pair of given points exchange order exactly once throughout the half-turn rotation.
    \item Hence, the number of inversions in the permutation $\sigma$ starts at $0$, increases at each combinatorial change, and ends up at $\binom{n}{2}$.
\end{itemize}

For each combinatorial change between $\sigma(\alpha-\varepsilon)$ and $\sigma(\alpha+\varepsilon)$ at critical angle $\alpha$, there corresponds a \emph{flip vector}
\begin{equation}\label{eq_flip_vector}
p_A(\alpha)=((\ell_0,r_0),\ldots,(\ell_{k-1},r_{k-1})),
\end{equation}
with $0\le\ell_0<r_0<\ell_1<\cdots<r_{k-1}< n$, which enumerates the left- and right-endpoints of the blocks that are reversed at critical angle $\alpha$.

The \emph{flip-vector sequence} $P_A=p_A(\alpha_0),\ldots,p_A(\alpha_{m-1})$, where $0<\alpha_0<\cdots<\alpha_{m-1}<\pi$ are the critical angles of $A$, encodes in a certain sense the combinatorial structure of $A$. If the set $B$ is a rotation of $A$, then the flip-vector sequence of $B$ will be of the form
\begin{equation}\label{eq_rot}
\begin{split}
P_B &= p_A(\alpha_i), \ldots, p_A(\alpha_{m-1}), \overline{p_A(\alpha_0)}, \ldots, \overline{p_A(\alpha_{i-1})}\qquad\text{or}\\
P_B &= \overline{p_A(\alpha_i)},\ldots,\overline{p_A(\alpha_{m-1})},p_A(\alpha_0),\ldots,p_A(\alpha_{i-1})
\end{split}
\end{equation}
for some $i$. Here $\overline{p_A(\alpha)}$ denotes the mirrored flip vector
\begin{equation*}
\overline{p_A(\alpha)} = ((n-1-r_{k-1},n-1-\ell_{k-1}),\ldots,(n-1-r_0,n-1-\ell_0))
\end{equation*}
for $p_A(\alpha)$ as in (\ref{eq_flip_vector}).

Similarly, if $B$ is a mirror image of $A$ under a vertical axis of symmetry, then
\begin{equation}\label{eq_mirror}
    P_B=p_A(\alpha_{m-1}),\ldots,p_A(\alpha_0).
\end{equation}

We can use this approach to find all $6$-point planar sets that have no singly-determined direction. If $A$ has no singly-determined direction, then no flip vector $p_A(\alpha)$ will be of the form $((\ell_1,\ell_1+1))$. A computer enumeration (see code in ancillary files) shows that there are $57$ possible ways to reach $(5,\ldots,0)$ starting from $(0,\ldots,5)$ and following the above-mentioned constraints: at each step inverting more than a single pair, and only inverting blocks that were in increasing order before the inversion.

\begin{table}
\begin{tabular}{rl}
0.& \texttt{(05)}\\
1.& \texttt{(01,23),(12,34),(01,23,45),(12,34),(23,45),(02,34)}\\
2.& \texttt{(12,34),(01,23,45),(12,34),(01,23,45),(12,34),(01,23,45)}\\
3.& \texttt{(13),(01,34),(23,45),(12,34),(01,23),(12,34),(01,45)}\\
4.& \texttt{(01,23),(12,34),(01,23,45),(12,34),(01,23),(12,45),(01,34)}\\
\end{tabular}
\par\vspace{1em}
\caption{\label{table_flip_vec_seqs}Flip-vector sequences produced by the enumeration program. For readability, we omitted inner parentheses and commas.}
\end{table}

\begin{table}
\begin{tabular}{rp{13cm}}
0.& \texttt{[abcdef] fedcba}\\
1.& \texttt{[ab][cd]ef b[ad][ce]f [bd][ae][cf] d[be][af]c de[bf][ac] [def][bc]a fedcba}\\
2.& \texttt{a[bc][de]f [ac][be][df] c[ae][bf]d [ce][af][bd] e[cf][ad]b [ef][cd][ab] fedcba}\\
3.& \texttt{a[bcd]ef [ad]c[be]f da[ce][bf] d[ae][cf]b [de][af]cb e[df][ac]b [ef]dc[ab] fedcba}\\
4.& \texttt{[ab][cd]ef b[ad][ce]f [bd][ae][cf] d[be][af]c [de][bf]ac e[df]b[ac] [ef]d[bc]a fedcba}
\end{tabular}
\par\vspace{1em}
\caption{\label{table_abcdef}A more convenient representation of the flip-vector sequences in Table \ref{table_flip_vec_seqs}.}
\end{table}

These $57$ cases correspond to $5$ possible flip-vector sequences $P_A$, up to the transformations in (\ref{eq_rot}) and (\ref{eq_mirror}). See Tables \ref{table_flip_vec_seqs} and \ref{table_abcdef}.

Sequence $0$ corresponds to a set of six points on a line.

\begin{figure}
    \centering
    \includegraphics[width=14cm]{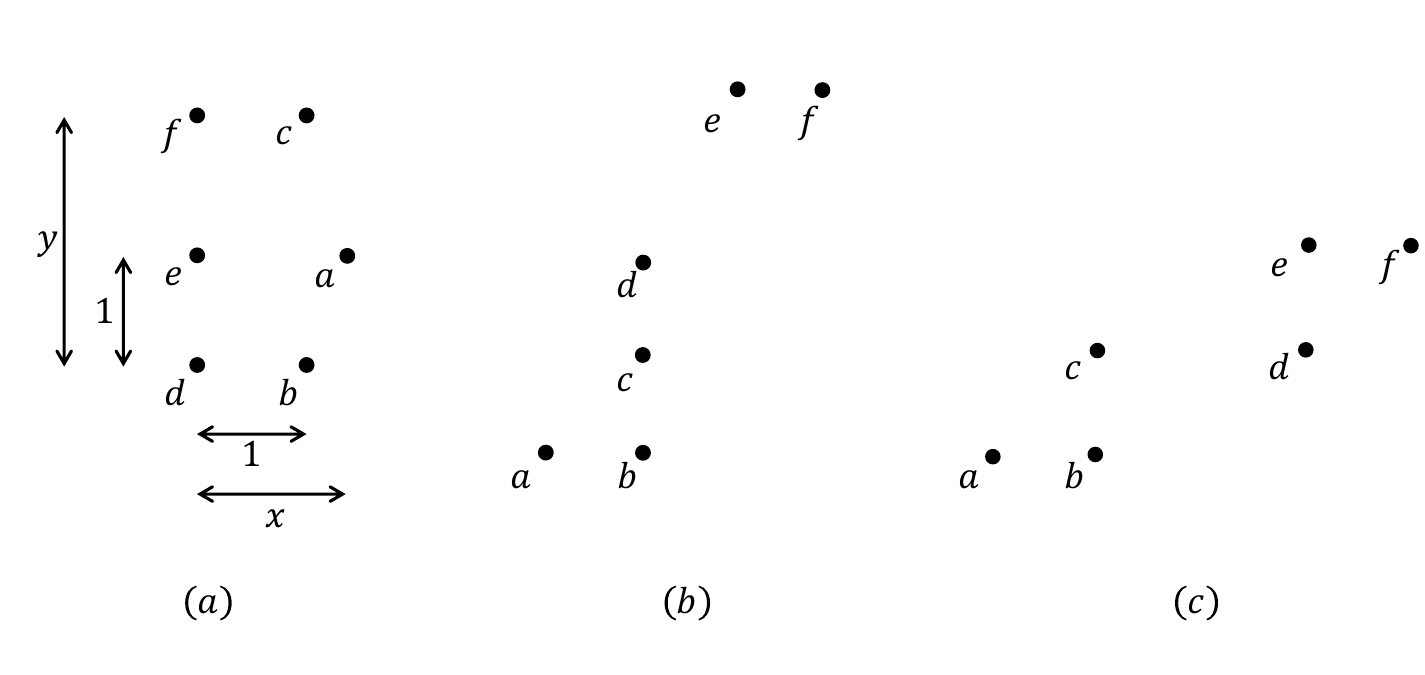}
    \caption{\label{fig_realize_flip}$(a)$ Proof that sequence $1$ is unrealizable. $(b)$ Proof that sequence $3$ is unrealizable. $(c)$ Proof that sequence $2$ can only be realized as an affine copy of a regular hexagon.}
\end{figure}

\begin{lemma}\label{lem_seq1}
Sequence $1$ is unrealizable.
\end{lemma}

\begin{proof}
    Without loss of generality let $b=(1,0)$, $d=(0,0)$, $e=(0,1)$ (as any three non-collinear points can be set arbitrarily by an affine transformation). Then relations \texttt{[def][bc]a} and \texttt{[bd][ae][cf]} imply $a=(x,1)$, $c=(1,y)$, $f=(0,y)$ for some $x>1$ and $y>1$. Then relation \texttt{d[be][af]c} implies $x=y-1$. But then $ad$ cannot be parallel to $ce$, contradicting relation \texttt{b[ad][ce]f}. See Figure \ref{fig_realize_flip}$(a)$.
\end{proof}

\begin{lemma}\label{lem_seq3}
Sequence $3$ is unrealizable.
\end{lemma}

\begin{proof}
    Without loss of generality let $a=(0,0)$, $b=(1,0)$, $c=(1,1)$. Then the relations \texttt{a[bcd]ef}, \texttt{[ef]dc[ab]}, \texttt{e[df][ac]b} imply $d=(1,y)$, $e=(x,z)$, $f=(z-y+1,z)$ for some $x$, $y$, $z$. Then the relations \texttt{[ad]c[be]f}, \texttt{da[ce][bf]}, \texttt{d[ae][cf]b}, \texttt{[de][af]cb} yield, respectively, the following equations:
    \begin{align*}
        y(x-1)&=z,\\
        (z-1)(z-y)&=z(x-1),\\
        z(z-y)&=(z-1)x,\\
        (z-y)(z-y+1)&=z(x-1).
    \end{align*}
    But then a software package yields that the only solution to these equations is $x=y=z=0$, which makes $e$ coincide with $a$. See Figure \ref{fig_realize_flip}$(b)$.
\end{proof}

\begin{lemma}
    Sequence $2$ can only be realized by an affine copy of a regular hexagon.
\end{lemma}

\begin{proof}
    Without loss of generality let $a=(0,0)$, $b=(1,0)$, $c=(1,1)$. Then relations \texttt{a[bc][de]f}, \texttt{[ef][cd][ab]}, \texttt{[ac][be][df]} yield $d=(x,1)$, $e=(x,x-1)$, $f=(2x-2, x-1)$ for some $x$. Then relation \texttt{c[ae][bf]d} yields $x=3$. See Figure \ref{fig_realize_flip}$(c)$.
\end{proof}

\begin{lemma}
    Sequence $4$ can only be realized by an affine copy of a regular heptagon without a vertex.
\end{lemma}

\begin{figure}
    \centering
    \includegraphics[width=15cm]{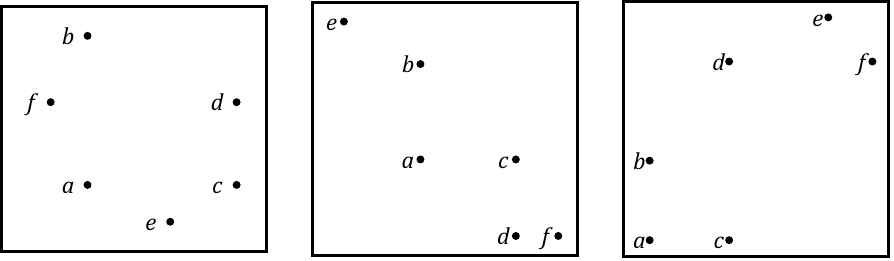}
    \caption{\label{fig_realize_hep}Realizing sequence $4$.}
\end{figure}

\begin{proof}
    It is easily verified that $G$ is a realization of sequence $4$, with $(a,b,c,d,e,f)=(p_5,p_4,p_6,p_3,p_2,p_1)$. We now show that sequence $4$ has a unique realization.

    Without loss of generality let $a=(0,0)$, $b=(0,1)$, $c=(1,0)$. Then relations \texttt{[ab][cd]ef}, \texttt{e[df]b[ac]}, \texttt{[ef]d[bc]a} yield $d=(1,y)$, $e=(z, y-z+x)$, $f=(x,y)$ for some $x, y, z$. Then relations \texttt{b[ad][ce]f}, \texttt{[bd][ae][cf]}, \texttt{d[be][af]c}, \texttt{[de][bf]ac } yield the equations
    \begin{align*}
        y(z-1)&=y-z+x,\\
        (y-1)z&=y-z+x,\\
        (y-1)(x-1)&=y,\\
        (y-z+x-1)x&=yz,\\
        (x-z)x&=(y-1)(z-1).
    \end{align*}
    A software package yields that this system has $3$ distinct solutions, in which $x$, $y$, and $z$ satisfy
    \begin{align*}
        x^3-4x^2+3x+1&=0,\\
        y^3-2y^2-y+1&=0,\\
        z^3-2z^2-z+1&=0.
    \end{align*}
    These solutions are plotted in Figure \ref{fig_realize_hep}. But only the rightmost solution respects the order of the points given by the relations.
\end{proof}

\end{document}